\documentclass[letterpaper, 11pt]{article}

\usepackage[T1]{fontenc}

\usepackage{geometry}
\usepackage{setspace}
\usepackage[style = phys, articletitle=true, backend=biber]{biblatex}
\usepackage{graphicx}
\usepackage{float}
\newfloat{scheme}{htbp}{los}
\floatname{scheme}{Scheme}
\floatname{chart}{Chart}
\newfloat{graph}{htbp}{loh}

\usepackage{chemformula} 

\usepackage{authblk}
\author[1]{Guillaume Lagüe*}
\author[1]{Frédérick Bernardot}
\author[2]{Mauricio Calvo}
\author[3]{Olfa Selmi}
\author[3]{Sofia Masi}
\author[1]{Christophe Testelin}
\author[2]{Hernan Miguez}
\author[3]{Ivan Mora-Sero}
\author[1]{Maria Chamarro}

\affil[1]{Sorbonne Université, CNRS, Institut des
NanoSciences, Paris 75005, France}
\affil[2]{Instituto de Ciencia de Materiales de Sevilla (ICMS)
Consejo Superior de Investigaciones Cientíﬁcas – Universidad de Sevilla (CSIC-US)
Américo Vespucio, 49, Sevilla 41092, Spain}
\affil[3]{Institute of Advanced Materials (INAM), Universitat Jaume I, Castelló de la Plana 12071, Spain}

\title{Hyperfine driven spin relaxation of charge carriers in metal-halide perovskites}
\date{*Email: guillaume.lague@insp.jussieu.fr}

\usepackage{comment}
\begin{document}

\maketitle

\newpage

\begin{abstract}
Spin relaxation of localized charge carriers in semiconductors is primarily governed by hyperfine interaction with the surrounding nuclear spin bath. While this mechanism is well-established in III–V bulk materials and quantum dots, its critical role in metal halide perovskites (MHPs) has only recently emerged. Their inverted band structure induces an unusual hierarchy of hyperfine couplings, with hole interactions dominating electron interactions, particurlaly in Pb-based perovskites.
Here, we adapt a spin relaxation model—originally developed for muon spin spectroscopy—to provide an exact description of longitudinal spin relaxation for localized carriers across arbitrary hyperfine correlation times. This approach is motivated by recent experimental evidence in MAPbI$_3$, which places carrier spins in an intermediate correlation regime, where conventional mono-exponential approximations fail.
Our analysis reveals distinct hyperfine relaxation channels: electrons couple primarily to halogen nuclei, whereas holes are governed by metal nuclei. This leads to a key prediction—perovskites with lighter halogens and metal cations exhibit significantly extended spin lifetimes. Applied to time-resolved Faraday rotation data obtained from two perovskite samples, our model extracts key microscopic parameters—including the carrier localization volume and hyperfine correlation time—demonstrating the necessity of the exact dynamical solution over a single-exponential approximation.
These findings provide microscopic insight into hyperfine-driven spin relaxation in MHPs and establish a robust framework for characterizing carrier localization and spin dynamics.

\end{abstract}

\section{I. Introduction}

The control of charge carrier spins in semiconductors is a central challenge for the development of spin-based technologies, including spintronics and quantum information processing \cite{SemiconSpintronics}\cite{OpticalOrientation}\cite{Spintronics}. The ability to initialize, manipulate, and preserve spin states over a long period of time is essential for these applications.  However, spin relaxation and decoherence remain major obstacles to the realization of scalable spin-based devices. In conventional semiconductors, spin relaxation of itinerant carriers at room temperature is dominated by spin–orbit coupling mechanisms such as the D’yakonov–Perel and Elliott Yafet processes, which lead to very short spin lifetimes \cite{OpticalOrientation}\cite{Spintronics}. These processes can be significantly suppressed at low temperature and in the presence of spatial confinement, where the restricted motion of charge carriers gives rise to alternative spin relaxation mechanisms.

In this latter regime, the dynamics of localized carrier spins is largely governed by the hyperfine interaction between carrier spins and the surrounding nuclear spin bath \cite{DPHyperfine}\cite{OpticalOrientationTheory}\cite{OpticalOrientationExperiment}. This mechanism has been extensively investigated in III–V bulk materials and semiconductor quantum dots, where the strong spatial confinement allows precise control of the carrier–nuclear spin interactions \cite{ESpinDecoherenceQDs}\cite{PRBMerkulov}\cite{NucSpinQD}\cite{TestelinHF}. Similar effects were also observed in quantum well semiconductors, where carriers localized on impurities or defect centers experience hyperfine-induced spin relaxation \cite{SpinRelaxGaAs}\cite{CdTeGuadalupe}.

Recently, metal halide perovskites have emerged as a new class of semiconductors exhibiting intriguing spin properties \cite{Even2013}\cite{Rashba1}\cite{Rashba2} in addition to their remarkable optoelectronic performance \cite{MHPfornextgen}\cite{NextgenOptical}. Evidences indicate that hyperfine interaction also plays a key role at low temperature in the spin relaxation of localized carriers in these systems \cite{CsPbBrBAYER}\cite{LeadDominatedHyperfine}\cite{MeliakovCsPbBr}\cite{FAPbBrKirstein}. A distinctive feature of lead-halide perovskites is their inverted band structure \cite{Even2012}, which results in unusual hyperfine couplings: hole spins interact more strongly with the nuclear spin bath than electron spins \cite{CsPbBrBAYER}\cite{LeadDominatedHyperfine}. This situation contrasts with most III–V semiconductors, where the hyperfine interaction is typically dominated by the conduction band electron \cite{TestelinHF}\cite{InAsDesfonds}. Furthermore, the specific orbital composition of the band edges in halide perovskites leads to a strongly anisotropic hyperfine interaction, giving rise to spin dynamics that differ significantly from those observed in conventional semiconductor systems \cite{NanoLettersLAGUE}.
 
Theoretical descriptions of hyperfine-induced spin relaxation in semiconductor quantum dots were developed assuming an infinite carrier lifetime in the localization potential, corresponding to an infinite hyperfine correlation time \cite{PRBMerkulov}\cite{TestelinHF}. However, experimental observations of spin dynamics in semiconductors require models that account for finite correlation times associated with carrier hopping, trapping, or recombination processes \cite{SpinHoppingGlazov}\cite{KudlacikFACsPbIBr}\cite{Shumilin2015}. In parallel, significant theoretical progresses have been achieved in other spin systems, such as muon spin relaxation \cite{kubo1967magnetic}\cite{DynamicKubo} and molecular spin dynamics \cite{SchultenWolynes} where more general frameworks for stochastic spin relaxation have been established. Advances in the theory of spin noise spectroscopy also gave rise to descriptions of effective carrier spin relaxation times in semiconductors \cite{SpinHoppingGlazov}\cite{SpinInertiaQD}\cite{Polarrecovery}.
 
 Here we demonstrate that a spin relaxation model originally developed for muon spin spectroscopy provides a powerful framework to describe the hyperfine-mediated relaxation of localized carriers in bulk perovskite semiconductors. We present an exact solution for the longitudinal spin relaxation in this context and investigate how the chemical composition of metal halide perovskites affects the relaxation dynamics. By applying this model to experimental spin relaxation data obtained on different perovskite samples, we extract key microscopic parameters, namely the carrier localization volume and the hyperfine correlation time. These results provide new insight into the role of hyperfine interaction in the spin dynamics of halide perovskites and highlight the potential of these materials for future spin-based technologies.

\section{II. Longitudinal spin relaxation of localized charge carriers in semiconductors}

\subsection{A. The Merkulov-Rosen-Efros relaxation function}

At low temperature, spin relaxation of localized charge carriers is primarily governed by the hyperfine interaction with lattice nuclei.
The hyperfine Hamiltonian describing the interaction between a carrier spin 
$\boldsymbol{S}$ and a nuclear spin $\boldsymbol{I}$ reads \cite{TestelinHF}\cite{Gryncharova}:

\begin{equation}
    H = 2 \mu_B \hbar\gamma_N \boldsymbol{I} \cdot \left\{\frac{8\pi}{3}\delta(r)\boldsymbol{S} + \frac{\boldsymbol{l}-\boldsymbol{S}}{r^3} + 3\frac{\left(\boldsymbol{S}\cdot\boldsymbol{r}\right)\boldsymbol{r}}{r^5}\right\} = g\,\mu_B\,\boldsymbol{S}\cdot\boldsymbol{B_N}(\boldsymbol{r})\,,
    \label{Hyperfine Hamiltonian}
\end{equation}

\smallskip
\noindent
where $\boldsymbol{l}$ is the orbital angular momentum of the electron or hole at the nucleus, $\boldsymbol{r}$ is the position vector of the charge carrier with respect to the nucleus, $\mu_B$ is the Bohr magneton, $\gamma_N$ the nuclear gyromagnetic ratio, $g$ is the carrier g-factor. Evaluating the matrix element of the hyperfine Hamiltonian allows one to define an effective nuclear magnetic field $\boldsymbol{B_N}$ \cite{PRBMerkulov}\cite{Abragam}\cite{TestelinHF}\cite{Gryncharova}.

This spin relaxation mechanism has been extensively studied in conventional semiconductors and is described by the Merkulov-Rosen-Efros model (MER) \cite{PRBMerkulov}\cite{SpinPhysicsinSC}, which is formally analogous to the Kubo-Toyabe model previously developed in muon spin relaxation theory \cite{kubo1967magnetic}\cite{PRBKubo}\cite{DynamicKubo}. In this framework, and in the absence of an applied magnetic field, the carrier spin precesses around the randomly oriented nuclear field $\boldsymbol{B_N}$. When an external magnetic field $\boldsymbol{B}$ is applied, the spin precesses around the total effective magnetic field $\boldsymbol{B_{tot}}=\boldsymbol{B}+\boldsymbol{B_N}$:

\begin{equation}
\frac{d\boldsymbol{S}}{dt} = \frac{g \mu_B (\boldsymbol{B}+\boldsymbol{B_N})}{\hbar} \times \boldsymbol{S}\,.
\end{equation}

\smallskip
Each localized carrier spin in the ensemble interacts with a different nuclear environment and therefore experiences a different nuclear field $\boldsymbol{B_N}$.
The fields $\boldsymbol{B_N}$ are distributed according to the normal distribution function \cite{PRBMerkulov}\cite{PRBKubo}:

\begin{equation}
P(\boldsymbol{B_N}) = \frac{1}{(\sqrt{2\pi}\Delta B_N)^3} \exp \left( -\frac{\boldsymbol{B_N^2}}{2\Delta B_N^2}\right),
\label{Distrib}
\end{equation}

\smallskip
\noindent
where $\Delta B_N$ characterizes the dispersion of the nuclear fields.

This model assumes a frozen nuclear spin bath, i.e. nuclear spin states are not modified by the interaction with carrier spins. This approximation is generally reasonable since the nuclear spin dynamics are much slower than the carrier spin dynamics \cite{PRBMerkulov}. Under these assumptions, an analytical expression for the relaxation of localized spins with zero external  magnetic field can be obtained \cite{PRBMerkulov}\cite{kubo1967magnetic}\cite{DynamicKubo}:

\begin{equation}
    \frac{\left<S_z\left(t\right)\right>}{S_{z,0}} = \frac{1}{3} + \frac{2}{3}\left(1-\left(\frac{t}{T_\Delta}\right)^2\right) \exp{\left(-\frac{1}{2} \left(\frac{t}{T_\Delta}\right)^2\right)}
    \label{Eq Merkulov B0}
\end{equation}

with $T_\Delta = \frac{1}{\gamma\Delta B_N}$ and $\gamma = \frac{g \mu_b}{\hbar}$ is the gyromagnetic ratio of the carrier spin. $S_{z}(t)$ is the longitudinal component of the spin $\boldsymbol{S}$ along the z-axis, and $S_{z,0}$ is its initial value. If a longitudinal magnetic field is applied such that $\Omega$ is the Larmor precession frequency of the considered spin in the applied field, the spin dynamics writes \cite{kubo1967magnetic}\cite{PRBKubo}\cite{DynamicKubo}:

\begin{equation}
    \frac{\left<S_z\left(t\right)\right>}{S_{z,0}} = 1-2\left(\frac{1}{\Omega T_\Delta}\right)^2 \left[ 1- \exp{\left(-\frac{1}{2}\left(\frac{t}{T_\Delta}\right)^2\right)} \cos{\Omega t}\right] + \frac{2}{\Omega ^3T_\Delta ^4} \int_{0}^{t} \exp{\left(-\frac{1}{2}\left(\frac{\tau}{T_\Delta}\right)^2\right) \sin{\Omega \tau}} d\tau
    \label{Eq Merkulov champ}
\end{equation}

\subsection{B. Finite hyperfine correlation time}

The model developed by Merkulov \textit{et al.} is strictly valid in the limit of an infinite correlation time, i.e., when the charge carrier remains indefinitely localized (e.g. in a defect or in a quantum dot). In realistic systems, the correlation time $\tau_c $ of localized charge carriers is finite and depends on temperature and defects density \cite{DPHyperfine}\cite{OpticalOrientationTheory}. When considering a carrier localized on a donor state, D'yakonov and Perel described the possibility for the carrier to jump from one donor site to another, effectively changing the nuclear field perceived by the carrier spin and therefore introducing a finite hyperfine correlation time \cite{DPHyperfine}\cite{OpticalOrientationTheory}. Experimental results in bulk Ga$_x$Al$_{1-x}$As, at that time, showed that this correlation time was relatively small compared to the longitudinal spin relaxation time ($\frac{\tau_c}{T_1} \ll 1$) \cite{OpticalOrientationExperiment}. A theory of longitudinal spin relaxation in the short correlation time regime was then established, giving the following expression of relaxation rate:

\begin{equation}
    T_1^{-1} = 2 \frac{\omega_N^2 \tau_c}{1+\left(\Omega \tau_c\right)^2},
    \label{T1shortcorr}
\end{equation}
 where $\omega_N = \frac{1}{T_{\Delta}} = \gamma \Delta B_N$ denotes the spin precession frequency in the fluctuating nuclear field.
 
 Later on, while assuming a Markovian spin dynamics and a mono-exponential spin relaxation, Smirnov \textit{et al.} defined an effective longitudinal spin relaxation time, for arbitrary hyperfine correlation time. This relaxation time is defined as \cite{SpinInertiaQD}:

\begin{equation}
    T_1 = \tau_c \frac{\left<\frac{1 + \Omega_{tot, z}^2\tau_c^2}{1 + \Omega_{tot}^2\tau_c^2}\right>}{1 - \left<\frac{1 + \Omega_{tot, z}^2\tau_c^2}{1 + \Omega_{tot}^2\tau_c^2}\right>}
\end{equation}

\noindent
where $\Omega_{tot} = \omega_N + \Omega$ is the carrier spin precession frequency in the sum of the nuclear and the applied fields, and $\left< . \right>$ denotes averaging over the nuclear spin fluctuations.
 
Lead halide perovskites constitute a notable case, as they often operate in the intermediate regime, $\frac{\tau_c}{T_1} \sim 1 $ ($\omega_N \tau_c \sim 1 $).

Being neither in the mono-exponential (short correlation time) or MER (infinite correlation time) regimes, we propose a generalization of both approaches, for metal halide perovskites, in order to consider any hyperfine correlation time. The effect of a finite correlation time is captured by the dynamical Kubo–Toyabe relaxation function in muon relaxation theory \cite{DynamicKubo}\cite{DTSCMDynamicKT}, and is similar to the framework developed by Schulten and Wolynes for studying electron spins in radicals \cite{SchultenWolynes}. In the following, we will denote this model the dynamical model. As described in \cite{DynamicKubo}\cite{SchultenWolynes}, the dynamical model extends the MER approach by introducing a finite probability that the nuclear field $\boldsymbol{B_N}$ is randomly changed to mimic carrier hopping. We give here a detailed development of this relaxation function inspired from Refs. \cite{DynamicKubo} and \cite{DTSCMDynamicKT}.
 
Each hopping event abruptly changes the local hyperfine field, with no correlation between successive fields; this approximation is known as the "strong collision" approximation \cite{SCMKubo1954}\cite{DynamicKubo}. To capture these dynamics, we consider the MER static relaxation function $g(t) = \frac{\left<S_z\left(t\right)\right>}{S_{z,0}}$ introduced in \textbf{Equation \ref{Eq Merkulov champ}}. The stochastic hopping of the charge carriers between localized sites occurs at a rate $\nu=\frac{1}{\tau_c}$. After hopping, the local field is randomly changed to a new value with a probability following the distribution function in \textbf{Equation \ref{Distrib}}. Each spin can experience a certain amount of hops, $n$, until time t. $g^{(n)}(t)$ is then the relaxation function for a spin experiencing exactly n hopping events between $t^\prime=0$ s and $t^\prime=t$ and can be written :
 
\begin{equation}
    g^{(0)}(t) = e^{-\nu t}g(t)
\end{equation}
\begin{equation}
    g^{(n)}(t) = \nu^n \int_0^t \int_0^{t_1} \ldots \int_0^{t_{n-1}} dt_1 \ldots dt_n e^{-\nu t}g(t-t_1)g(t_1-t_{2}) \ldots g(t_{n-1}-t_{n})g(t_n) \ \ (n\ge1)
\end{equation}
\noindent 
We then define the general relaxation function of a localized charge carrier spin as the sum of the relaxation functions $g^{(n)}(t)$:
\begin{equation}
G(t)=\sum_{n=0}^{+\infty} g^{(n)}(t).
\end{equation}
\noindent
For an infinite correlation time, ($\nu=0$), $g^{(n)}(t) = 0$ for $ n \ge 1$, and we recover the expected limit:  $G(t)=g^{(0)}(t) = g(t)$. This relaxation function $G(t)$ can be numerically solved using Laplace transform methods \cite{DynamicKubo}\cite{DTSCMDynamicKT}. However, for better computation efficiency, we used in this work the discrete time strong collision model (DTSCM) of the dynamic Kubo-Toyabe relaxation function as developed in Ref. \cite{DTSCMDynamicKT}.

\begin{figure}[H]
    \centering
    \includegraphics[width=0.70\linewidth]{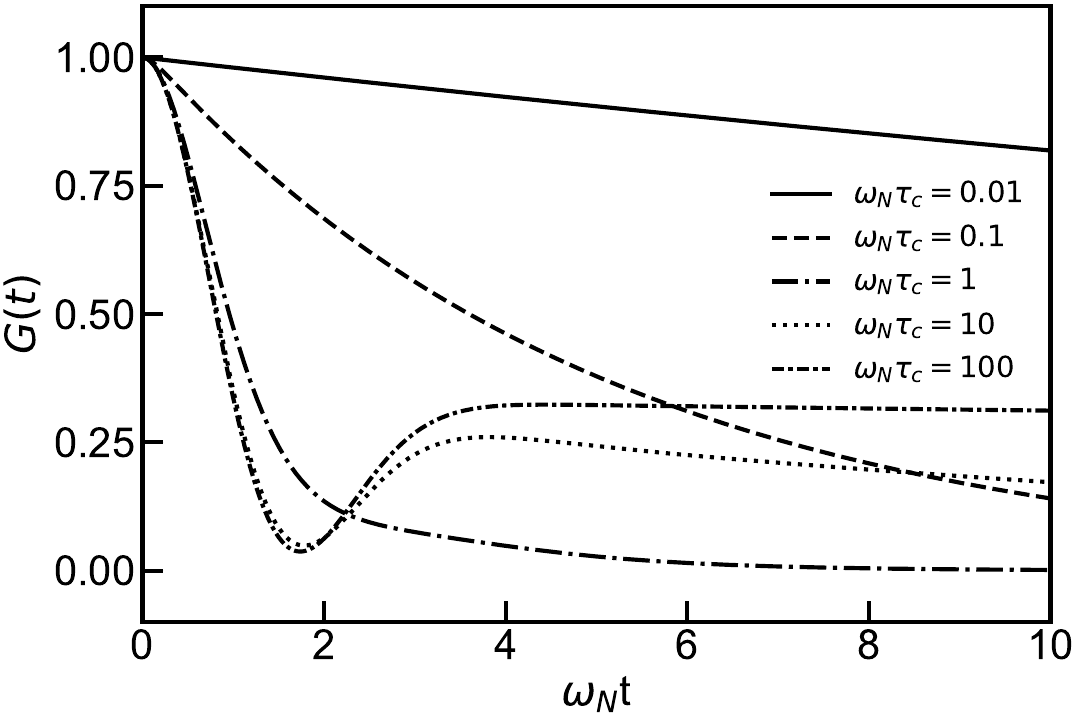}
    \caption{Longitudinal spin relaxation function computed using the dynamical Kubo-Toyabe model presented in this work for correlation times ranging from $\omega_N \tau_c = 0.01$ to $\omega_N \tau_c = 100$}
    \label{Figure 1}
\end{figure}

The relaxation functions computed using te dynamical model, with different correlation times and no external magnetic field are displayed in \textbf{Figure \ref{Figure 1}}. As the considered correlation time evolves, three distinct regimes can be identified. In the short correlation time regime, illustrated by $\omega_N \tau_c = 0.01$, we are in the domain of motional narrowing. In this regime, rapid fluctuations of the hyperfine field effectively average out the inhomogeneous broadening, thereby reducing dephasing and extending the effective spin relaxation time. We therefore observe a very slow longitudinal spin relaxation. As the correlation time increases, the system enters an intermediate regime where spin relaxation between successive hopping events becomes increasingly efficient. In the range $\omega_N \tau_c = 0.01$ to $\omega_N \tau_c = 1$, the relaxation rate steadily increases. Finally for large correlation times ($\omega_N \tau_c \geq 10$), the system approaches the infinite correlation approximation described by the MER model, characterized by a fast initial Gaussian-like decay followed by a long-living plateau at one-third of the initial spin polarization. 
These results, presented in \textbf{Figure \ref{Figure 1}}, demonstrate that the treatment of hyperfine correlation times is crucial to accurately model the spin relaxation of localized charge carriers in semiconductors.

\begin{figure}[H]
    \centering
    \includegraphics[width=0.70\linewidth]{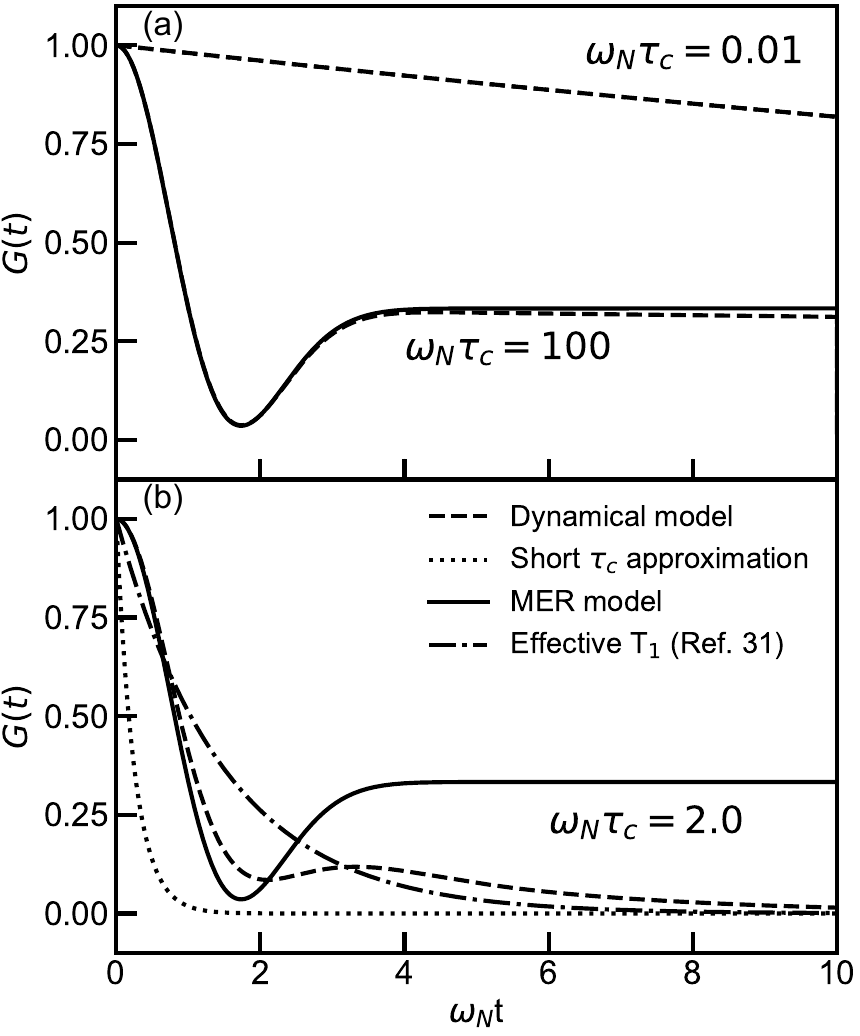}
    \caption{Comparison between the three longitudinal spin relaxation functions presented in this work: the dynamical Kubo-Toyabe function (dashed line), the short $\tau_c$ approximation (dotted line) and the MER model (solid line), as a function of reduced time $\omega_N t$. (a) The dynamical Kubo-Toyabe and the short $\tau_c$ approximation functions, for $\omega_N \tau_c$=0.01. At this scale, both curves are indiscernable, showing the remarkably good agreement between the two functions. The dynamical model and MER functions are also given for $\omega_N \tau_c$=100. Very similar behaviors are observed. (b) Relaxation functions for the three models computed with $\omega_N \tau_c$=2.0}
    \label{Figure 2}
\end{figure}

A comparison of the exact solution of the longitudinal spin relaxation function, introduced by the model used in this work, with the MER model and the short correlation time approximation is given in \textbf{Figure \ref{Figure 2}}. This figure highlights the ability of the dynamical model to describe the full range of correlation times.
 
In particular, \textbf{Figure \ref{Figure 2} (a)} shows that this model successfully reproduces the behaviors of both the MER model and the short correlation time approximation, corresponding respectively to long and short correlation time regimes. However, \textbf{Figure \ref{Figure 2} (b)} illustrates the stark difference between the three approaches in the intermediate regime, where $\omega_N \tau_c \sim 1$. In this regime, each model predicts a distinct relaxation function. Neither the short $\tau_c$ approximation nor the MER model provides an accurate description of the longitudinal spin relaxation of localized charge carrier spins. Furthermore, although we see a better agreement, the effective mono-exponential $T_1$ approach, supposedly valid for all values of correlation times \cite{SpinInertiaQD}, displays a different spin relaxation function than the dynamical Kubo-Toyabe function. The difference lies in the assumptions made when developing an effective longitudinal relaxation time $T_1$. This effective relaxation time implies a mono-exponential relaxation function. However, as shown in \textbf{Figure \ref{Figure 2} (b)}, the exact solution of the Kubo-Toyabe model is clearly non-exponential. These observations underline the necessity of introducing an exact treatment, given here by the dynamical Kubo–Toyabe function, to properly capture the spin dynamics in this intermediate regime.

\subsection{C. Hyperfine interaction quenching by a longitudinal magnetic field}

\begin{figure}[H]
    \centering
    \begin{frame}{}
    \includegraphics[width=0.95\linewidth]{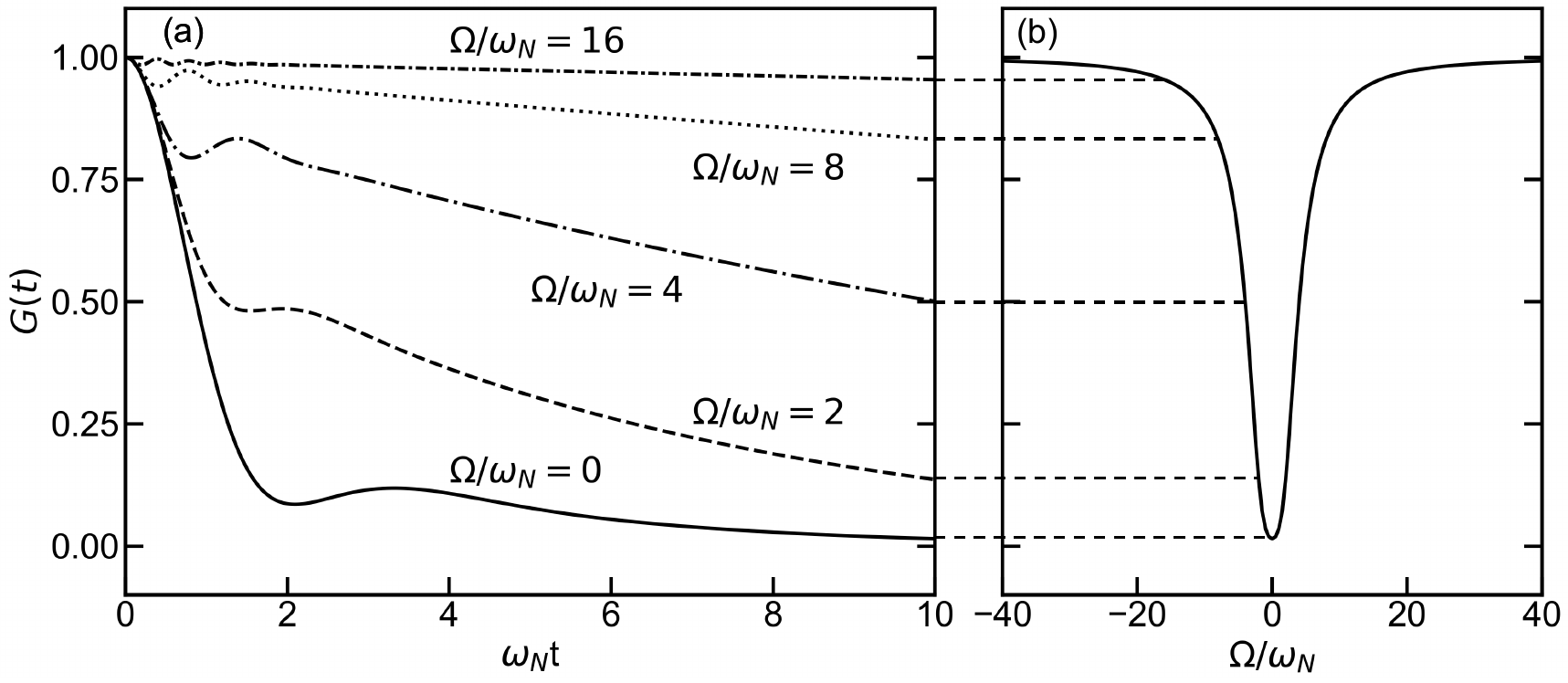}
    \caption{(a) Longitudinal spin relaxation function as a function of reduced time $\omega_N t$ at different longitudinal magnetic fields, computed with $\omega_N \tau_c = 2.0$; (b) longitudinal component of carrier spin G(T), taken at $\omega_N T = 10$, as a function of reduced longitudinal magnetic field $\Omega / \omega_N$.}
    \label{Figure 3}
    \end{frame}
\end{figure}

The application of a longitudinal magnetic field leads to the decrease of the relaxation rate via hyperfine interaction \cite{DPHyperfine}\cite{OpticalOrientationTheory}\cite{PRBMerkulov}, as illustrated by \textbf{Equations (\ref{Eq Merkulov champ})} and \textbf{(\ref{T1shortcorr})}. \textbf{Figure \ref{Figure 3} (a)} displays the evolution of the longitudinal spin relaxation function $G(t)$ as the amplitude of the magnetic field is increased, considering the dynamical model in the intermediate regime ($\omega_N \tau_c = 2.0$). Experimentally, this evolution can also be illustrated by observing how the projection of the longitudinal component of a carrier spin $G(t)$, at a fixed time T, evolves with an applied magnetic field. As illustrated in \textbf{Figure \ref{Figure 3} (b)}, we observe an increase of the longitudinal spin component as the magnetic field increases.

\section{III. Hyperfine induced spin relaxation of electrons and holes in metal halide perovskites}

\subsection{A. Conduction and valence band Bloch states in lead halide perovskites}

The hyperfine coupling of charge-carrier spins in halide perovskites is highly sensitive to two factors: the nuclear species within the lattice and the charge-carrier wavefunctions. From the nuclear perspective, different species exhibit different hyperfine coupling strengths. On the charge carriers side, the strength of the hyperfine interactions depends on the orbital character of the Bloch states: carriers with a non-zero density of probability at the  nucleus (e.g. s-type orbitals) are subject to a contact interaction, whereas those with zero density at the nucleus (e.g. purely p-type orbitals) interact only through the weaker dipole-dipole mechanism.
Density  functional theory (DFT) calculations for halide perovskites can be used to compute the orbital contributions of various atoms at the valence band maximum and the conduction band minimum. In particular, the organic cation plays a minor role in the electronic states near the band gap, whereas the inorganic Pb-X framework (where X is a halogen) is the dominant contribution \cite{NanoLettersLAGUE}. The Bloch functions in a cubic crystal then take the following form:

\begin{equation}
\left|vb_{+\frac{1}{2}}\right> = \left[ C_{Pb,6s}\left|S_0\right> + \frac{C_{hal,5p}}{\sqrt{3}}\left(\left|P_{X, 1}\right> + \left|P_{Y, 2}\right> + \left|P_{Z, 3}\right>\right) \right] \left| \uparrow \right>
\label{vb+}
\end{equation}
\begin{equation}
\left|vb_{-\frac{1}{2}}\right> = \left[ C_{Pb,6s}\left|S_0\right> + \frac{C_{hal,5p}}{\sqrt{3}}\left(\left|P_{X, 1}\right> + \left|P_{Y, 2}\right> + \left|P_{Z, 3}\right>\right) \right] \left| \downarrow \right>
\end{equation}
\noindent
for holes in the valence band, and:

\begin{equation}
    \begin{split}
\left|cb_{+\frac{1}{2}}\right> = & \frac{1}{\sqrt{3}} \left[C_{hal,5s}\left|S_3\right> + C_{Pb,6p}\left|P_{Z, 0}\right>\right] \left| \uparrow \right> \\
&- \frac{1}{\sqrt{3}} \left[ C_{Pb,6p}\left|P_{X, 0}\right> + C_{hal,5s}\left|S_1\right> + i \left( C_{Pb,6p}\left|P_{Y, 0}\right> + C_{hal,5s}\left|S_2\right> \right)\right] \left| \downarrow \right> 
    \end{split}
\end{equation}
\begin{equation}
    \begin{split}
\left|cb_{-\frac{1}{2}}\right> = & \frac{1}{\sqrt{3}} \left[C_{hal,5s}\left|S_3\right> + C_{Pb,6p}\left|P_{Z, 0}\right>\right] \left| \downarrow \right> \\
&- \frac{1}{\sqrt{3}} \left[ C_{Pb,6p}\left|P_{X, 0}\right> + C_{hal,5s}\left|S_1\right> - i \left( C_{Pb,6p}\left|P_{Y, 0}\right> + C_{hal,5s}\left|S_2\right> \right)\right] \left| \uparrow \right> 
    \end{split}
    \label{cb-}
\end{equation}
\noindent
for electrons in the conduction band.
$\left|S\right>$, $\left|P_X\right>$, $\left|P_Y\right>$ and $\left|P_Z\right>$ represent the atomic s and p-type orbitals. The nucleus indexed 0 is the lead nucleus located in the middle of the unit cell and the nuclei indexed 1, 2 and 3 represent the three different halogen atoms as schematized in \textbf{Figure \ref{Figure 4}}. DFT calculations performed on several halide perovskites give $0.2 < \left|C_{Pb, 6s}\right|^2 < 0.4$ in the valence band, and $0.80 < \left|C_{Pb, 6p}\right|^2 < 0.95$ in the conduction band \cite{DFTNature}\cite{DFTexpLee}\cite{DFTMAPI}\cite{DFTNatComm}\cite{PDOSCsPbBr}\cite{PDOSCsPbI}\cite{PDOSFAPI}. We then deduce the admixture of the halogen orbitals as $\left|C_{Pb}\right|^2 + \left|C_{hal}\right|^2 = 1$ in both the top of the valence band and the bottom of the conduction band.\\

\subsection{B. Nuclear magnetic field distributions}

As detailed earlier in \textbf{Equation (\ref{Hyperfine Hamiltonian})}, the hyperfine Hamiltonian can be expressed as an effective nuclear field acting on carrier spins. The nuclear magnetic field acting on an electron spin is then given by \cite{DNP2026BAYER} \cite{NanoLettersLAGUE}:

\begin{equation}
    \boldsymbol{B_{N, e}} = \frac{v_0}{Z\mu_Bg_e} \sum_{i=1}^N \left( \sum_{j=1}^Z \sum_{k=1, 2, 3} |\psi_e\left(R_{hal_k}^{i_j} \right)|^2 A_{hal_{k}}^{cb} \boldsymbol{I_{hal_k}^{i_j}} 
+ \sum_{j=1}^Z \eta_{Pb} |\psi_e\left(R_{Pb}^{i_j} \right)|^2 A_{Pb}^{cb} \boldsymbol{I_{Pb}^{i_j}} \right),
\label{BNe}
\end{equation}
\noindent
where the sum is carried over the N unit cells within the localization volume of the carrier. $v_0$ is the unit cell volume, and Z denotes the number of unit formulas within a unit cell ($Z=$ 1, 2 or 4 depending on the crystal structure of the perovskite). $\boldsymbol{I_{Pb}}$, $\boldsymbol{I_{hal_1}}$, $\boldsymbol{I_{hal_2}}$ and $\boldsymbol{I_{hal_3}}$ are the nuclear spin operators of the lead and the three non-equivalent halogen nuclei \cite{NanoLettersLAGUE}\cite{DNP2026BAYER}, respectively.  $\psi_e\left(R\right)$  denotes the envelope function of the conduction band electron at the position R. $A_{hal_{1}}^{cb}$, $A_{hal_{2}}^{cb}$, $A_{hal_{3}}^{cb}$ and $A_{Pb}^{cb}$ are the effective hyperfine coupling constants of an electron in the conduction band with the three non-equivalent halogen nuclei in the perovskite lattice and the $^{207}$Pb nuclei, respectively. We give in \textbf{Appendix A} the expressions for these hyperfine coupling matrices.

In \textbf{Equation \ref{BNe}}, we define $\eta_k \in \{0,1\}, \  \langle \eta_k \rangle = \alpha_{Pb}$
to consider the natural abundance $\alpha_{Pb}$ = 0.22 of the $^{207}$Pb isotope which is the only naturally occurring lead isotope with non zero spin.

The variance $\Delta B_N^{\ 2} = \frac{\langle B_N^{\ 2} \rangle}{3}$ of the nuclear field distribution is then defined as:

\begin{equation}
    \Delta B_{N, e}^2 = \frac{Z}{3}\left(\frac{v_0}{Z\mu_Bg_e}\right)^2 \sum_{i=1}^N |\psi_e(R_i)|^4 \left\{ \alpha_{Pb} (A_{Pb}^{cb})^2I_{Pb}(I_{Pb}+1) + \left((A_{hal_{1}}^{cb})^2 + (A_{hal_{2}}^{cb})^2 + (A_{hal_{3}}^{cb})^2\right)I_{hal}(I_{hal}+1)\right\}.
    \label{DBNe}
\end{equation}

A similar treatment for hole spins gives:

\begin{equation}
    \boldsymbol{B_{N, h}} = \frac{v_0}{Z\mu_Bg_h} \sum_{i=1}^N \left( \sum_{j=1}^Z \sum_{k=1, 2, 3} |\psi_h\left(R_{hal_k}^{i_j} \right)|^2 A_{hal_{k}}^{vb} \boldsymbol{I_{hal_k}^{i_j}} 
+ \sum_{j=1}^Z \eta_{Pb} |\psi_h\left(R_{Pb}^{i_j} \right)|^2 A_{Pb}^{vb} \boldsymbol{I_{Pb}^{i_j}} \right),
\label{BNh}
\end{equation}

and the variance of the nuclear field distribution is given by:

\begin{equation}
    \Delta B_{N, h}^2 = \frac{Z}{3}\left(\frac{v_0}{Z\mu_Bg_h}\right)^2 \sum_{i=1}^N |\psi_h(R_i)|^4 \left\{ \alpha_{Pb} (A_{Pb}^{vb})^2I_{Pb}(I_{Pb}+1) + \left((A_{hal_{1}}^{vb})^2 + (A_{hal_{2}}^{vb})^2 + (A_{hal_{3}}^{vb})^2\right) I_{hal}(I_{hal}+1)\right\}.
    \label{DBNh}
\end{equation}

Here again, we give an explicit expression for the effective hyperfine coupling constant $A_{Pb}^{vb}$ and $A_{hal_1}^{vb}$, $A_{hal_2}^{vb}$, $A_{hal_3}^{vb}$ in \textbf{Appendix A}.

The sums in \textbf{Equations (\ref{DBNe})} and \textbf{(\ref{DBNh})} run over all nuclei placed within the localization volume of the charge carriers. For a donor-like localized state of radius $a_0$, the envelope function can be written as  $\psi(r) = \frac{1}{\sqrt{\pi a_0^3}}e^{\frac{-r}{a_0}}$. We can therefore write:

\begin{equation}
    \Delta B_{N, e}^2 = \frac{1}{3}\frac{v_0}{Z\mu_B^2 g_e^2} \frac{1}{8\pi a_0^3}\left\{ \alpha_{Pb} (A_{Pb}^{cb})^2I_{Pb}(I_{Pb}+1) + \left((A_{hal_{1}}^{cb})^2 + (A_{hal_{2}}^{cb})^2 + (A_{hal_{3}}^{cb})^2\right) I_{hal}(I_{hal}+1)\right\}
    \label{DBNe_integrated}
\end{equation}

and

\begin{equation}
    \Delta B_{N, h}^2 = \frac{1}{3}\frac{v_0}{Z\mu_B^2 g_h^2} \frac{1}{8\pi a_0^3}\left\{ \alpha_{Pb} (A_{Pb}^{vb})^2I_{Pb}(I_{Pb}+1) + \left((A_{hal_{1}}^{vb})^2 + (A_{hal_{2}}^{vb})^2 + (A_{hal_{3}}^{vb})^2\right) I_{hal}(I_{hal}+1) \right\},
    \label{DBNh_integrated}
\end{equation}

\subsection{C. Dependence of the longitudinal spin relaxation function on the halogen composition}

Because the orbitals of A$^+$ cations in halide perovskite materials do not participate to the Bloch states, as illustrated by \textbf{Equations (\ref{DBNe_integrated})} and \textbf{(\ref{DBNh_integrated})}, changes of  cation (A$^+$) have very little effect on the longitudinal spin relaxation of electrons and holes. Strictly speaking, changing the cation will modify the volume of the unit cell and eventually the crystal structure of the perovskite but no change in the hyperfine couplings will be observed.

However, modifying the halogen in metal halide perovskite has a direct influence on hyperfine couplings of localized carriers and will therefore strongly modify their spin relaxation. We give in \textbf{Tables \ref{table eff hf holes}} and \textbf{\ref{table eff hf electrons}} the effective hyperfine coupling constants between electrons (holes) in the conduction (valence) band and lead as well as different halogen atoms, as a function of the lead admixture coefficients to the Bloch states. These values were obtain from theoretical calculations of atomic coupling constants \cite{HFcouplingconstants}. We underline here that these hyperfine coupling constants were computed from Roothaan-Hartree-Fock atomic wavefunctions \cite{CLEMENTIROETTI} and are slightly smaller than the theoretical constants determined using the Hartree-Fock-Slater atomic orbitals computed by Herman and Skillman \cite{HermanSkillman}\cite{MortonPreston}. Experimentally determined hyperfine coupling constants of these atoms display a large spread, depending on both the material studied and the experimental technique used \cite{NanoLettersLAGUE}\cite{DNP2026BAYER}. We note a difference between the constants adopted in the present work and those from previous studies of hyperfine interaction in lead halide perovskites \cite{DNP2026BAYER}. For Pb s orbitals, for which the largest body of experimental data exist, the constant obtained from Ref. \cite{HFcouplingconstants} (used in this work) falls within the experimental reported range, while the constant obtained with Ref. \cite{MortonPreston} (used in ref. \cite{DNP2026BAYER}) lies on the higher edge of the results. In the absence of experimental results for bulk lead halide perovskites, we decided to use the theoretical values given by Ref. \cite{HFcouplingconstants}.

\begin{table}[H]
    \centering
    \begin{tabular}{c c c c c}

        & $A^{vb}_{Pb}$ & $A^{vb}_{I}$ & $A^{vb}_{Br}$ & $A^{vb}_{Cl}$ \\
       \hline
       \hline
      $\left|C_{Pb, 6s}\right|^2$ = 0.20 & 48 $\mu eV$ & 0.69 $\mu eV$ & 0.71 $\mu eV$ & 0.16 $\mu eV$ \\
      \hline
      $\left|C_{Pb, 6s}\right|^2$ = 0.40 & 96 $\mu eV$ & 0.52 $\mu eV$ & 0.54 $\mu eV$ & 0.12 $\mu eV$ \\

    \end{tabular}
    \caption{Effective hyperfine coupling constants between holes in the valence band and the nuclei in the perovskite lattice.}
    
    \label{table eff hf holes}
\end{table}

\begin{table}[H]
    \centering
    \begin{tabular}{c c c c c}
        & $A^{cb}_{Pb}$ & $A^{cb}_{I}$ & $A^{cb}_{Br}$ & $A^{cb}_{Cl}$ \\
       \hline
       \hline
      $\left|C_{Pb, 6p}\right|^2$ = 0.80 & 11 $\mu eV$ & 8.8 $\mu eV$ & 6.9 $\mu eV$ & 1.3 $\mu eV$ \\
      \hline
      $\left|C_{Pb, 6p}\right|^2$ = 0.95 & 13 $\mu eV$ & 2.2 $\mu eV$ & 1.7 $\mu eV$ & 0.33 $\mu eV$ \\

    \end{tabular}
    \caption{Effective hyperfine coupling constants between electrons in the conduction band and the nuclei in the perovskite lattice.}
    
    \label{table eff hf electrons}
\end{table}

We propose in this section to study the carrier spin relaxation dependence on  the halide composition of the perovskite material. We consider three different perovskites: MAPbI$_3$, MAPbBr$_3$ and MAPbCl$_3$.  MAPbBr$_3$ and MAPbCl$_3$ crystals are cubic at room temperature, while MAPbI$_3$ is tetragonal. We will however consider it cubic for the sake of the comparison. Their respective unit cell volumes are $v_{0, I} = 252$ \AA$^3$, $v_{0, Br} = 209$ \AA$^3$ and $v_{0, Cl} = 184$ \AA$^3$ \cite{UnitCellVolumes}. This work is still applicable to non-cubic crystals, provided that the Bloch states defined in \textbf{Equations (\ref{vb+})} to \textbf{(\ref{cb-})} will be slightly modified and so will the hyperfine coupling matrices given in \textbf{Appendix A}. The three crystals display one formula unit per unit cell ($Z = 1$) \cite{UnitCellVolumes}. Furthermore, we consider a localization radius of carriers $a_0 = 5$ nm for both carriers in all three perovskites. The hyperfine coupling constants are computed considering $\left|C_{Pb, 6s}\right|^2 = 0.30$ and $\left|C_{Pb, 6p}\right|^2 = 0.88$, according to the values reported in \textbf{Tables \ref{table eff hf holes}} and \textbf{\ref{table eff hf electrons}}. We also consider the value of the three halogen spins I$_I$ = $\frac{5}{2}$, I$_{Br}$ = $\frac{3}{2}$, I$_{Cl}$ = $\frac{3}{2}$ along with the spin of $^{207}$Pb, I$_{Pb}$ = $\frac{1}{2}$, and a hyperfine correlation time $\tau_c = 4.0$ ns.

\begin{figure}[H]
    \centering
    \includegraphics[width=0.70\linewidth]{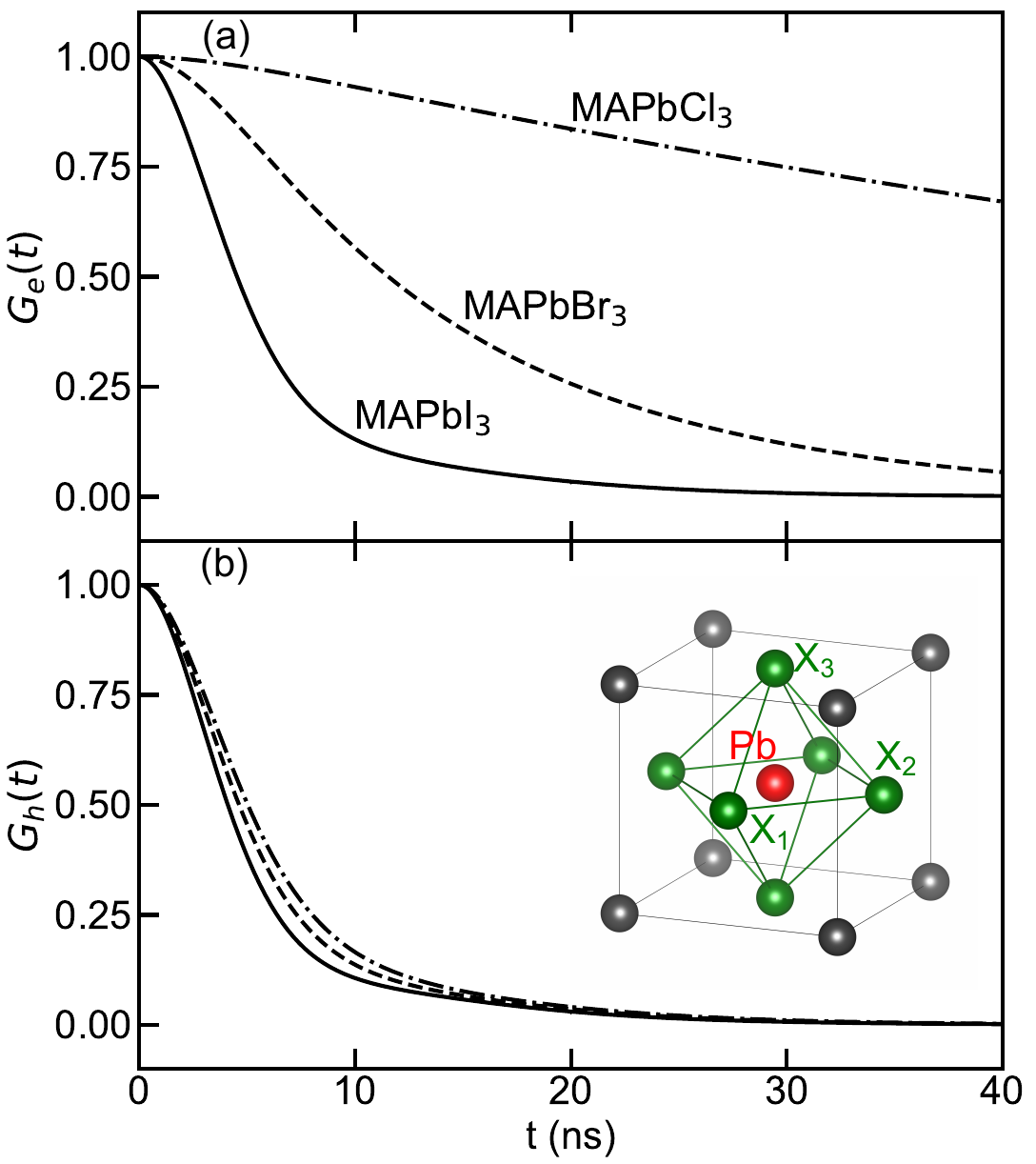}
    \caption{Longitudinal carrier spin relaxation function in the three perovskites considered: MAPbI$_3$ (solid line), MAPbBr$_3$ (dashed line) and MAPbCl$_3$ (dot-dashed line). (a) Electron spin relaxation function, showing an increase in the spin relaxation time as the halogen in the perovskite gets smaller; (b) Hole spin relaxation, showing no strong dependence on the halogen species. Inset: representation of the perovskite crystal MAPbX$_3$ (X = I, Br, Cl), showing the three different halogen atoms X$_{i}$ ($i=1,2,3$).}
    \label{Figure 4}
\end{figure}

As illustrated in \textbf{Figure \ref{Figure 4} (a)}, electron spin relaxation is highly dependent on the halogen composition of the perovskite. The smaller the halogen nucleus, the weaker the hyperfine interaction with the electron and therefore the slower the longitudinal spin relaxation. We obtain electron spin relaxation times ranging from several nanoseconds for MAPbI$_3$ up to hundreds of nanoseconds for MAPbCl$_3$. This strong reduction of the relaxation rate (sensitive to the hyperfine coupling between electron and halogen) is induced by the simultaneous decreases of unit cell volume, halogen nucleus spin and hyperfine coupling constant.\\
Hole spin relaxation, however, is not highly dependent on the halogen species. As evidenced in earlier works \cite{NanoLettersLAGUE}\cite{LeadDominatedHyperfine}\cite{DNP2026BAYER}, hyperfine interaction with holes is mainly governed by lead nuclei. The slight differences between hole spin relaxation rates displayed in \textbf{Figure \ref{Figure 4} (b)} are mainly due to the changes in unit cell volume between the three perovskites. All three perovskites present longitudinal hole spin relaxation times of several nanoseconds, in agreement with experimental findings \cite{SpinCoherenceLAGUE}\cite{CsPbBrBAYER}\cite{LeadDominatedHyperfine}\cite{JPCLGuadalupe}.\\

A similar analysis can be applied to a modification of the heavy atom in the perovskite, which provides the largest contribution to the hyperfine interaction. One way to achieve this is by replacing lead with tin. We will therefore now compare carrier spin relaxation in two different perovskites: MAPbI$_3$ and MASnI$_3$. Once again we will consider a cubic structure, with volumes $v_{0, Pb} = 252$ \AA$^3$ and $v_{0, Sn} = 242$ \AA$^3$ for MAPbI$_3$ and MASnI$_3$ respectively \cite{UnitcellStoumpos}. We further assume that substituting lead atoms with tin in the perovskite does not significantly modify the Bloch wavefunctions, and we suppose that the orbital admixture coefficient remain unchanged upon substitution.\\

We first need to determine the hyperfine coupling constants describing the interaction between electrons in the conduction band as well as holes in the valence band with tin nuclei in tin based perovskites. These constants are given in \textbf{Tables \ref{table eff hf Sn holes}} and \textbf{\ref{table eff hf Sn electrons}} along with the hyperfine constants with lead nuclei in lead based perovskites, taken from Ref. \cite{HFcouplingconstants}.

\begin{table}[H]
    \centering
    \begin{tabular}{c c c}

        & $A^{vb}_{Pb}$ & $A^{vb}_{Sn}$\\
       \hline
       \hline
      $\left|C_{Pb, 6s / Sn, 5s}\right|^2$ = 0.20 & 48 $\mu eV$ & 8.6 $\mu eV$ \\
      \hline
      $\left|C_{Pb, 6s / Sn, 5s}\right|^2$ = 0.40 & 96 $\mu eV$ & 17.2 $\mu eV$ \\

    \end{tabular}
    \caption{Effective hyperfine coupling constants between holes in the valence band and the nuclei in the perovskite lattice.}
    
    \label{table eff hf Sn holes}
\end{table}

\begin{table}[H]
    \centering
    \begin{tabular}{c c c}
        & $A^{cb}_{Pb}$ & $A^{cb}_{Sn}$ \\
       \hline
       \hline
      $\left|C_{Pb, 6p / Sn, 5p}\right|^2$ = 0.80 & 11 $\mu eV$ & 4.0 $\mu eV$ \\
      \hline
      $\left|C_{Pb, 6p / Sn, 5p}\right|^2$ = 0.95 & 13 $\mu eV$ & 4.8 $\mu eV$ \\

    \end{tabular}
    \caption{Effective hyperfine coupling constants between electrons in the conduction band and the nuclei in the perovskite lattice.}
    
    \label{table eff hf Sn electrons}
\end{table}

\begin{figure}[H]
    \centering
    \includegraphics[width=0.70\linewidth]{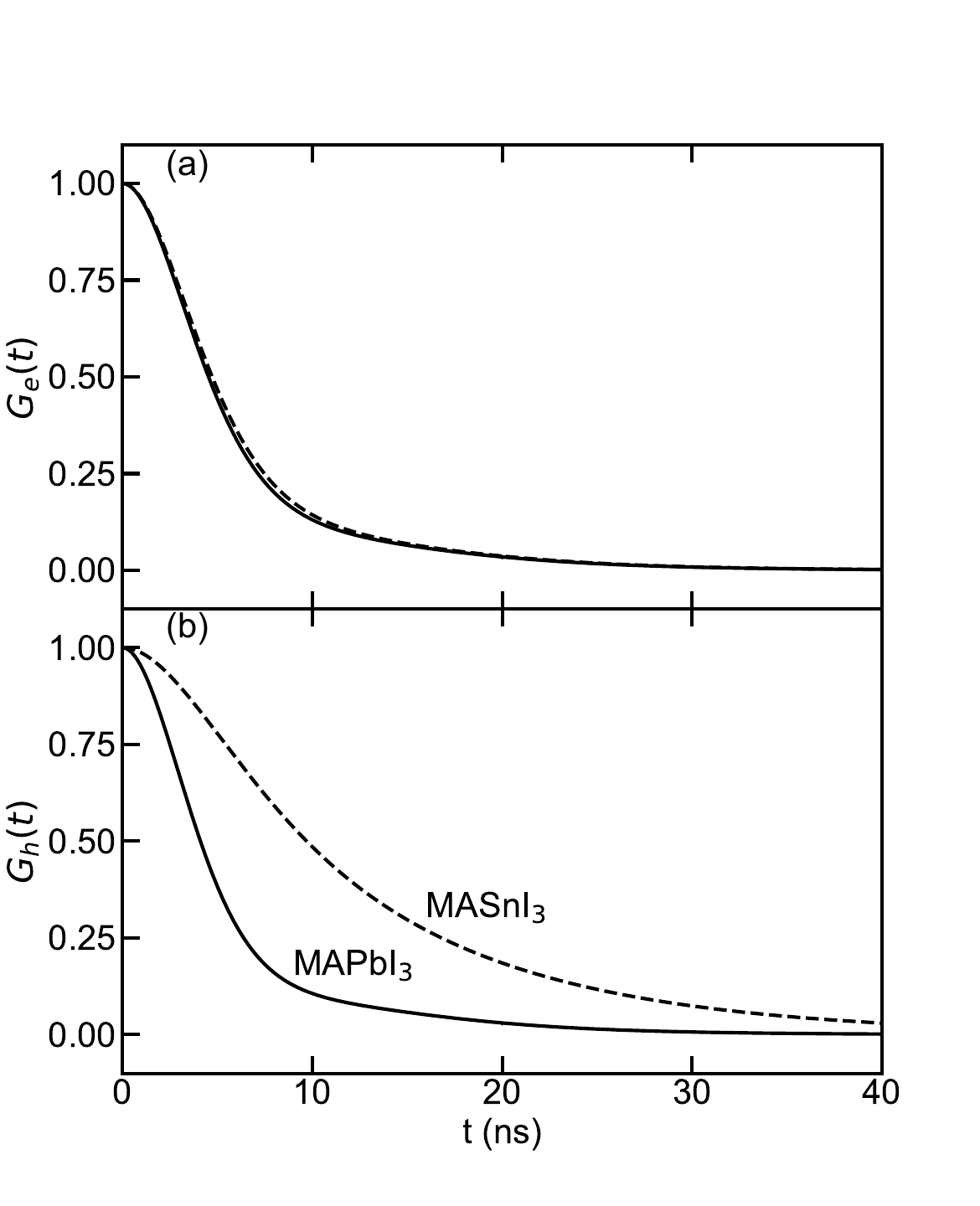}
    \caption{Longitudinal carrier spin relaxation function for the two perovskites considered: MAPbI$_3$ (solid line), MASnI$_3$ (dashed line). (a) Electron spin relaxation function, showing no significant difference between the two perovskites; (b) Hole spin relaxation, showing a slower longitudinal spin relaxation for MaSnI$_3$ than for MAPbI$_3$.}
    \label{Relaxation Pb et Sn}
\end{figure}

The first observation is that tin-related hyperfine interaction is significantly weaker than the lead-related interaction. We thus expect the hyperfine-induced spin relaxation to be less efficient, leading to longer spin relaxation times. Furthermore, as for lead the natural isotopic abundance $\alpha_{Pb}=0.226$ needs to be considered. Tin also presents a majority of naturally occurring isotopes with zero nuclear spins. We therefore define the parameter $\alpha_{Sn} = 0.163$ corresponding to the combined abundance of $^{117}$Sn and $^{119}$Sn isotopes, both of which have nuclear spin $\frac{1}{2}$. Using the same parameters as in the previous, namely carrier localization radii of 5 nm and hyperfine correlation times of 4,0 ns, we compute the longitudinal spin relaxation functions of both carriers in MAPbI$_3$ and MASnI$_3$. The resulting relaxation functions are displayed in \textbf{Figure \ref{Relaxation Pb et Sn}}. While replacing lead with tin has only a minor effect on the electron spin relaxation, it drastically modifies the hole spin relaxation. Similarly to what we observe for halogen substitution, the smaller the nucleus (Sn compared to Pb), the slower the longitudinal spin relaxation. For identical localization parameters, the hole spin relaxation time in MASnI$_3$ is predicted to be approximately twice as long as in MAPbI$_3$.

\section{IV. Experimental study of longitudinal spin relaxation in MAPbI$_3$ and FAPbI$_3$ thin films: localization radii and hyperfine correlation times}

\begin{figure}[H]
    \centering
    \includegraphics[width=0.90\linewidth]{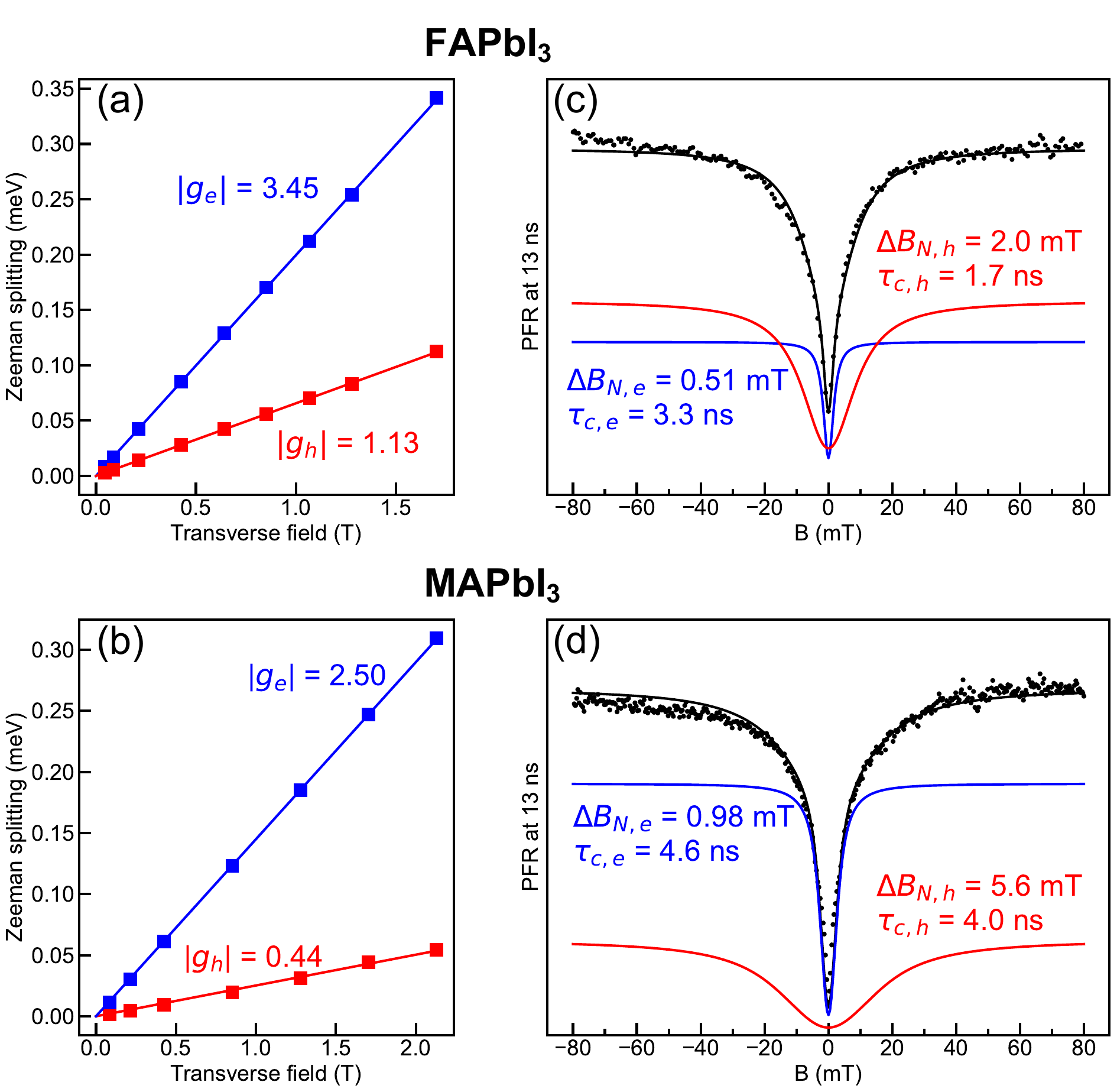}
    \caption{Zeeman splitting of carrier spins in a transverse magnetic field in (a) the FAPI sample and (b) the MAPI sample. PFR signal measured in a longitudinal magnetic at a pump-probe delay of 13 ns in (c) FAPI and (d) MAPI. These measurements were performed at 2 K with a laser set at 1.48 eV for FAPI and 1.64 eV for MAPI.}
    \label{Figure 5}
\end{figure}

In this section, we present experimental results of pump-probe photo-induced Faraday rotation (PFR) measurements performed on MAPbI$_3$ (MAPI) and FAPbI$_3$ (FAPI) polycrystalline samples, extracting electron and hole localization volumes as well as hyperfine correlation times. Details on sample preparations are given in \textbf{Appendix B} for MAPI thin films and \textbf{Appendix C} for FAPI thin films. We used a 2 ps Ti:Saphirre laser with repetition rate 76 MHz to perform the measurements. The excitation energy was set at the band gap energy of the studied sample (1.48 eV for FAPI and 1.64 eV for MAPI). The samples, placed in a 2 K cryostat, are excited by a circularly polarized pump pulse, creating a spin polarization in the sample \cite{SpinCoherenceLAGUE}\cite{Odenthalquantumbeatings}. A linearly-polarized pulse is then used to probe the spin polarization within the sample via PFR measurements. Applying a transverse magnetic field (Voigt configuration) allows to observe the precession of the carrier spins \cite{SpinCoherenceLAGUE}\cite{JPCLGuadalupe}, and therefore leads to the determination of g-factors, as illustrated in \textbf{Figure \ref{Figure 5} (a)} and \textbf{(b)}. As demonstrated in previous works, in FAPI the electron g-factor is $g_e = 3.45$, while the hole g-factor is $g_h = -1.13$ \cite{SpinCoherenceLAGUE}. In MAPI carrier g-factors measurements give $g_e = 2.50$ and $g_h = -0.44$ for electrons and holes respectively \cite{NanoLettersLAGUE}. To study hyperfine relaxation in these materials however, we work in a Faraday configuration (longitudinal magnetic field). The pump-probe delay was set at 13 ns, such that only localized carrier spins can be attributed to the PFR signal \cite{NanoLettersLAGUE}. A longitudinal magnetic field is then applied, leading to a rise in the PFR signal, as described in \textbf{Figure \ref{Figure 3} (b)}. We then use the dynamical model introduced earlier to fit the resulting curves, keeping in mind that both electron and hole spins contribute to the PFR signal. \textbf{Figure \ref{Figure 5} (c)} and \textbf{(d)} display the results of these measurements.

We obtain for the FAPI sample $\Delta B_{N, e} = 0.51$ mT and $\Delta B_{N, h} = 2.0$ mT as well as correlation times $\tau_{c, e} = 3.3$ ns and $\tau_{c, h} = 1.7$ ns, for electron and hole, respectively. As for the MAPI sample, we measured $\Delta B_{N, e} = 0.98$ mT and $\Delta B_{N, h} = 5.6$ mT along with $\tau_{c, e} = 4.6$ ns and $\tau_{c, h} = 4.0$ ns.

In bulk halide perovskites the longitudinal relaxation time $T_1$ in absence of an applied magnetic field, is usually measured in the range of several nanoseconds \cite{SpinCoherenceLAGUE}\cite{JPCLGuadalupe}\cite{LeadDominatedHyperfine}. Here, we demonstrate that bulk halide perovskite fall in the intermediate regime $\frac{\tau_c}{T_1} \sim 1$. Properly describing the longitudinal spin relaxation of charge carriers in these materials therefore requires the use of the dynamical model introduced in this work. Indeed, we previously reported the use of the short $\tau_c$ approximation to describe the PFR measured in MAPI, giving correlation times approximately half as long as those obtained with the exact solution provided by the Kubo-Toyabe model \cite{NanoLettersLAGUE}. Therefore, although other spin relaxation models can fit the experimental data in lead halide perovskites, one needs to be very cautious when interpreting the parameters extracted from such fits.

The full width at half maximum (FWHM) of the PFR signal obtained in these two samples is comparable to that derived from the photoluminescence studies under continuous excitation in FA$_{0.9}$Cs$_{0.1}$PbI$_{2.8}$Br$_{0.2}$ and MAPbI$_3$, using Lorentzian fits \cite{DNP2026BAYER}\cite{PRC2026Kotur}.
However, we must stress out that, as illustrated in \textbf{Figures \ref{Figure 5} (b)} and \textbf{(d)}, the FWHM of the fitting curves is significantly larger than the standard deviation $\Delta B_N$ of the corresponding nuclear field distribution, computed using the model developed in this work. A comprehensive analysis of the longitudinal spin relaxation, such as that performed here or introduced in Ref. \cite{Polarrecovery} for continuous wave measurements, is therefore necessary to correctly determine $\Delta B_N$ for each carrier.
Using \textbf{Equations (\ref{DBNe_integrated})} and \textbf{(\ref{DBNh_integrated})}, the value of $\Delta B_N$ allows us to determine the localization radius of each charge carrier in both samples. The crystallographic parameters of the two perovskites considered are reported in \textbf{Table \ref{Table crystal params}}. We note here that the crystal structures considered are the stable phases at 2 K of the FAPI and MAPI. We thus consider a tetragonal phase for FAPI and an orthorhombic phase for MAPI.
We must also emphasize that the Bloch states in \textbf{Equations (\ref{vb+})} to \textbf{(\ref{cb-})}, as well as the hyperfine coupling matrices developed in \textbf{Appendix A}, are given for a cubic structure. Bloch states of a perovskite in a tetragonal phase are given in Refs. \cite{Guadalupekp}\cite{NanoLettersLAGUE}. Switching from a cubic structure to a tetragonal structure and going further to an orthorhombic structure will slightly modify the Bloch states. These changes will induce a symmetry breaking between the crystallographic axes. However, we do not expect a strong modification in the hyperfine interaction between localized carrier spins and nuclei.

\begin{table}[H]
    \centering
    \begin{tabular}{c c c c c}
        
         & $v_0$ (\AA$^3$) & Crystal structure & Space group & Z\\
         \hline
         \hline
        FAPbI$_3$ & 494$^a$ & Tetragonal & P4/mbm & 2 \\
        MAPbI$_3$ & 950$^b$ & Orthorhombic & Pnma & 4 \\
        
    \end{tabular}
    \caption{Table of the crystallographic parameters of FAPI and MAPI.}
    \label{Table crystal params}
    
\end{table}
\noindent
${}^a$ Ref. \cite{FAPILattice}; ${}^b$ Ref. \cite{MAPILattice}\\

The computed localizations are reported in \textbf{Table \ref{localization MAPI FAPI}}. From the fitting of the experimental data, we can obtain the Larmor precession frequency $\omega_{N, e/h} = \gamma_{e/h}\Delta B_N$ of the charge carrier spins in the random nuclear fields. Similarly as we did for the theoretical relaxation functions computed in the previous part and shown in \textbf{Figure \ref{Figure 4}}, we considered $\left|C_{Pb, 6s}\right|^2 = 0.30$ and $\left|C_{Pb, 6p}\right|^2 = 0.88$.
Finally we obtained localization radii in FAPI of $a_{0, e} = 6.2$ nm for electrons and $a_{0, h} = 5.6$ nm for holes. In MAPI, we computed $a_{0, e} = 4.9$ nm for electrons and $a_{0, h} = 5.1$ nm for holes. These results are in agreement with previous reports on other lead halide perovskites \cite{CsPbBrBAYER}\cite{LeadDominatedHyperfine}.

\begin{table}[H]
    \centering
    \begin{tabular}{c c c c|c c c}
        
         & $g_e$ & $\omega_{N, e}$ (ns$^{-1}$) & $a_{0, e}$ (nm) & $g_h$ & $\omega_{N, h}$ (ns$^{-1}$) & $a_{0, h}$ (nm) \\
         \hline
         \hline
         FAPbI$_3$ & 3.45 & 0.15 & 6.2 & -1.13 & 0.20 & 5.6 \\
         MAPbI$_3$ & 2.50 & 0.22 & 4.9 & -0.44 & 0.22 & 5.1 \\
         
    \end{tabular}
    \caption{Table of the carriers g-factors, precession frequency in the random nuclear fields and localization radii in polycrystalline FAPI and MAPI.}
    \label{localization MAPI FAPI}
    
\end{table}
The extracted hyperfine correlation times, which fall within the nanosecond regime, align closely with the relatively slow spin dynamics observed in both FAPbI$_3$ and MAPbI$_3$. These experimental results offer strong empirical validation of the use of the Kubo dynamical model to describe spin relaxation in lead halide perovskites. 
Additionally, the estimated carrier localization radii—on the order of 5 nm—suggest a weak localization regime. Collectively, these findings bridge microscopic spin interactions with the spatial extent of charge carriers, thereby corroborating the robustness of the theoretical framework.

\section{V. Conclusion}

We develop a comprehensive model for the longitudinal spin relaxation of localized charge carriers in semiconductors, adapted from muon spin relaxation theory. While simple analytical expressions capture the limiting cases of short and long correlation times, the intermediate regime—highly relevant to real materials—requires an exact treatment. Here, we provide such a description through the dynamical Kubo–Toyabe formalism, valid for arbitrary correlation times.

We apply this framework to hyperfine-induced spin relaxation of localized electrons and holes in lead halide perovskites, a class of materials characterized by unconventional hyperfine interactions. We demonstrate that electron spin relaxation is strongly governed by the halogen species, with lighter nuclei yielding markedly extended spin lifetimes. In particular, we predict hyperfine-limited electron spin lifetimes reaching hundreds of nanoseconds in chlorine-based perovskites. We also evidence the influence of the metallic cation on hole spin relaxation, leading to a spin relaxation in tin-based perovskites twice as slow as in lead-based perovskites for a given localization volume.

We further support our model with experimental investigations of polycrystalline MAPbI$_3$ and FAPbI$_3$. From these measurements, we extract electron hyperfine correlation times   $\tau_{c, e} = 3.3$ ns and $\tau_{c, h} = 1.7$ ns in FAPI and $\tau_{c, e} = 4.6$ ns and $\tau_{c, h} = 4.0$ ns in MAPI. Quantitative analysis also provides direct estimates of carrier localization radii, giving $a_{0, e} = 6.2$ nm and $a_{0, h} = 5.6$ nm in FAPI and $a_{0, e} = 4.9$ nm and $a_{0, h} = 5.1$ nm in MAPI.

All together, these results deepen the understanding of hyperfine-driven spin relaxation in halide perovskites, revealing them as a promising platform for long-lived spin coherence. They also demonstrate the ability to link spin-related measurements to important microscopic parameters like carrier localization volumes.

\section{Acknowledgments}
This work is partially funded by the Ministerio de Ciencia e Innovación of the Spanish Government by the project PLEDs, PID2022-140090OB-C21/AEI/10.13039/501100011033/FEDER. O.S. also acknowledges Santiago Grisolia program under project CIGRIS/2022/122 funded by Generalitat Valenciana, and also by the French National Research Agency (ANR SPOIR, ANR-24-CE50-2579)

\section{Appendix A}

Hyperfine coupling matrices in lead halide perovskites can be determined using the basis of the Bloch states in \textbf{Equations (\ref{vb+})} to \textbf{(\ref{cb-})}, similarly as Refs. \cite{TestelinHF} and \cite{Gryncharova}.

In the valence band, the hyperfine coupling matrices are given as \cite{DNP2026BAYER}\cite{NanoLettersLAGUE}:

\begin{equation}
    A_{Pb}^{vb}\, =\, \left|C_{Pb, 6s}\right|^2 \frac{16\pi}{3} \hbar \mu_B \gamma_{Pb} |\psi(0)|^2
\end{equation}

\begin{equation}
    A^{vb}_{hal_1}\, =\, \frac{\left|C_{hal, 5p}\right|^2}{3} \frac{4}{5} \hbar \mu_B \gamma_I \left<\frac{1}{r^3}\right>
\begin{pmatrix}
    2 & 0 & 0\\
    0 & -1 & 0\\
    0 & 0 & -1
\end{pmatrix}\,,
\end{equation}

\begin{equation}
    A^{vb}_{hal_2}\, =\, \frac{\left|C_{hal, 5p}\right|^2}{3} \frac{4}{5} \hbar \mu_B \gamma_I \left<\frac{1}{r^3}\right>
\begin{pmatrix}
    -1 & 0 & 0\\
    0 & 2 & 0\\
    0 & 0 & -1
\end{pmatrix}\,,
\end{equation}

\begin{equation}
    A^{vb}_{hal_3}\, =\, \frac{\left|C_{hal, 5p}\right|^2}{3} \frac{4}{5} \hbar \mu_B \gamma_I \left<\frac{1}{r^3}\right>
\begin{pmatrix}
    -1 & 0 & 0\\
    0 & -1 & 0\\
    0 & 0 & 2
\end{pmatrix}\,,
\end{equation}
where
\begin{equation}
\left<\frac{1}{r^3}\right> = \int_\Omega dr X(r)^2 \frac{1}{r^3}\,,
\end{equation}
and $X(r)$ is the $P_x$ atomic orbital function.

And in the conduction band, we have \cite{DNP2026BAYER}\cite{NanoLettersLAGUE}:

\begin{equation}
    A_{Pb}^{cb}\, =\, \left|C_{Pb, 6p}\right|^2 \hbar \mu_B \gamma_{Pb}\frac{16}{3}\left<\frac{1}{r^3}\right> \,,
\end{equation}

\begin{equation}
A_{I_{1}}^{cb}\, =\, \frac{\left|C_{I, 5s}\right|^2}{3} \frac{16\pi}{3} \hbar \mu_B \gamma_I |\psi(0)|^2
\begin{pmatrix}
    1 & 0 & 0\\
    0 & -1 & 0\\
    0 & 0 & -1
\end{pmatrix},
\end{equation}

\begin{equation}
A_{I_{2}}^{cb}\, =\, \frac{\left|C_{I, 5s}\right|^2}{3} \frac{16\pi}{3} \hbar \mu_B \gamma_I |\psi(0)|^2
\begin{pmatrix}
    -1 & 0 & 0\\
    0 & 1 & 0\\
    0 & 0 & -1
\end{pmatrix},
\end{equation}

\begin{equation}
A_{I_{3}}^{cb}\, =\, \frac{\left|C_{I, 5s}\right|^2}{3} \frac{16\pi}{3} \hbar \mu_B \gamma_I |\psi(0)|^2
\begin{pmatrix}
    -1 & 0 & 0\\
    0 & -1 & 0\\
    0 & 0 & 1
\end{pmatrix},
\end{equation}

\section{Appendix B}
The polycrystalline MAPbI$_3$ film was fabricated via spin coating using a precursor solution obtained by combining PbI$_2$ and methylammonium iodide (MAI) in a 1:1 molar proportion. An appropriate amount of a 4:1 (v/v) mixture of N,N-dimethylformamide (DMF) and dimethyl sulfoxide (DMSO) was added to reach a final total concentration of 30\% wt. The film deposition was carried out using a single-step spin-coating process, in which the glass substrate with solution on top was spun at 3000 rpm for 60 seconds. In the course of the spinning process, a quantity of 230 $\mu$L of chlorobenzene was deposited onto the film surface after a period of 20 seconds.

\section{Appendix C}

All materials had high purity and were used as received without additional purification steps. Formamidinium iodide (FAI, 98\% purity, GreatCell Solar) and lead iodide (PbI$_2$, >98\% purity, TCI) were used. The solvents used were N,N-dimethylformamide (DMF, anhydrous, 99.9\% purity), dimethyl sulfoxide (DMSO, anhydrous, 99.9\% purity), and chlorobenzene (CB, anhydrous, 99.8\% purity), all purchased from Sigma-Aldrich.

To prepare the FAPbI$_3$ precursor solution, 645.41 mg (1.4 M) of PbI$_2$ and 240.75 mg of FAI (1.4 M) were dissolved in 1 mL of a DMF/DMSO mixed solvent (4:1, v/v). The solution was then stirred and heated at 60 $^\circ$C for 4 hours. Prior to use, it was filtered through a 0.22 $\mu$m filter to remove any undissolved particles and impurities, ensuring high solution purity and uniformity for deposition.

Glass substrates were sequentially cleaned in an ultrasonic bath using acetone, ethanol, and isopropanol for 10 minutes each, then dried with an N$_2$ flow, followed by a 5-minute UV/ozone treatment. The substrates were then introduced into a nitrogen-filled glove box for the deposition of the FAPbI$_3$ layer. The perovskite solution was deposited onto the glass substrates by spin-coating at 4000 rpm for 30 seconds. Then, 100 $\mu$L of chlorobenzene was dropped onto the substrate after 25 seconds of spinning to remove residual solvent and promote the formation of a well-organized crystalline layer. After the deposition of the perovskite solution, the substrates were immediately annealed at 100 $^\circ$C for 10 minutes to enhance crystallization and achieve a high-quality crystal structure.

\printbibliography

@book{SemiconSpintronics,
author = {Awschalom, D. and Loss, Daniel and Samarth, Nitin},
year = {2002},
month = {01},
pages = {},
title = {Semiconductor Spintronics and Quantum Computation},
isbn = {978-3-642-07577-3},
doi = {10.1007/978-3-662-05003-3}
}

@article{MHPfornextgen,
    author = {Dong, H and Ran, C.X and Gao, W.Y and Li, M.J and Xia, Y.D and Huang, W},
    title = {Metal Halide perovskite for next-generation optoelectronics:progresses and Prospects},
    journal = {eLight},
    year = 2023
}

@article{NextgenOptical,
author = {Quan, Li Na and Rand, Barry P. and Friend, Richard H. and Mhaisalkar, Subodh Gautam and Lee, Tae-Woo and Sargent, Edward H.},
title = {Perovskites for Next-Generation Optical Sources},
journal = {Chemical Reviews},
volume = {119},
number = {12},

year = {2019},
doi = {10.1021/acs.chemrev.9b00107},
    note ={PMID: 31021609}
}

@article{Even2013,
author = {Even, Jacky and Pedesseau, Laurent and Jancu, Jean-Marc and Katan, Claudine},
title = {Importance of Spin–Orbit Coupling in Hybrid Organic/Inorganic Perovskites for Photovoltaic Applications},
journal = {The Journal of Physical Chemistry Letters},
volume = {4},
number = {17},

year = {2013},
doi = {10.1021/jz401532q},

}

@Article{Rashba1,
author ="Liu, Beichen and Gao, Huaxiong and Meng, Chaoying and Ye, Honggang",
title  ="The Rashba effect in two-dimensional hybrid perovskites: the impacts of halogens and surface ligands",
journal  ="Phys. Chem. Chem. Phys.",
year  ="2022",
volume  ="24",
issue  ="45",

doi  ="10.1039/D2CP03971K",
}

@article{Rashba2,
author = {Ghosh, Supriya and Pradhan, Bapi and Bandyopadhyay, Arkamita and Skvortsova, Irina and Zhang, Yiyue and Sternemann, Christian and Paulus, Michael and Bals, Sara and Hofkens, Johan and Karki, Khadga J. and Materny, Arnulf},
title = {Rashba-Type Band Splitting Effect in 2D (PEA)2PbI4 Perovskites and Its Impact on Exciton–Phonon Coupling},
journal = {The Journal of Physical Chemistry Letters},
volume = {15},
number = {31},

year = {2024},
doi = {10.1021/acs.jpclett.4c01957},
}

@article{Even2012,
  title = {Electronic model for self-assembled hybrid organic/perovskite semiconductors: Reverse band edge electronic states ordering and spin-orbit coupling},
  author = {Even, J. and Pedesseau, L. and Dupertuis, M.-A. and Jancu, J.-M. and Katan, C.},
  journal = {Phys. Rev. B},
  volume = {86},
  issue = {20},

  year = {2012},

  doi = {10.1103/PhysRevB.86.205301},
}

@article{JPCLGuadalupe,
author = {Garcia-Arellano, Guadalupe and Tripp{\'e}-Allard, Gaëlle and Legrand, Laurent and Barisien, Thierry and Garrot, Damien and Deleporte, Emmanuelle and Bernardot, Fr{\'e}d{\'e}rick and Testelin, Christophe and Chamarro, Maria},
title = {Energy Tuning of Electronic Spin Coherent Evolution in Methylammonium Lead Iodide Perovskites},
journal = {The Journal of Physical Chemistry Letters},
volume = {12},
number = {34},

year = {2021},
doi = {10.1021/acs.jpclett.1c01790},

}

@article{SpinCoherenceLAGUE,
author = {Lag{\"u}e, Guillaume and Bernardot, Frederick and Guilloux, Victor and Legrand, Laurent and Barisien, Thierry and Sánchez-Diaz, Jesús and Galve-Lahoz, Sergio and Saïdi, Imen and Boujdaria, Kais and P. Martinez-Pastor, Juan and Testelin, Christophe and Mora-Seró, Iván and Chamarro, Maria},
title = {Spin Coherence and Relaxation Dynamics of Localized Electrons and Holes in FAPbI3 Films},
journal = {ACS Photonics},
volume = {11},
number = {7},
year = {2024},
doi = {10.1021/acsphotonics.4c00632},

}

@article{Guadalupekp,
  title = {Land\'e $g$ factors in tetragonal halide perovskite: A multiband $\mathrm{k}.\mathrm{p}$ model},
  author = {Garcia-Arellano, G. and Boujdaria, K. and Chamarro, M. and Testelin, C.},
  journal = {Phys. Rev. B},
  volume = {106},
  issue = {16},

  year = {2022},

  doi = {10.1103/PhysRevB.106.165201},

}

@article{Odenthalquantumbeatings,
author = {Odenthal, P. and Talmadge, W. and Gundlach, N. et al.},
title = {Spin-polarized exciton quantum beating in hybrid organic–inorganic perovskites},
journal = {Nature Phys},
volume = {13},

year = {2017},
doi = {10.1038/nphys4145},

}

@article{CsPbBrBAYER,
  title = {Coherent spin dynamics of electrons and holes in CsPbBr3 perovskite crystals},
  author = {Belykh, V.V. and Yakovlev, D.R. and Glazov, M.M. et al.},
  journal = {Nat Commun},
  volume = {10},
  
  year = {2019},

  doi = {10.1038/s41467-019-08625-z},
}

@article{LeadDominatedHyperfine,
author = {Kirstein, Erik and Yakovlev, Dmitri R. and Glazov, Mikhail M. and Evers, Eiko and Zhukov, Evgeny A. and Belykh, Vasilii V. and Kopteva, Nataliia E. and Kudlacik, Dennis and Nazarenko, Olga and Dirin, Dmitry N. and Kovalenko, Maksym V. and Bayer, Manfred},
title = {Lead-Dominated Hyperfine Interaction Impacting the Carrier Spin Dynamics in Halide Perovskites},
journal = {Advanced Materials},
volume = {34},
number = {1},

doi = {https://doi.org/10.1002/adma.202105263},

year = {2022}
}

@article{MeliakovCsPbBr,
  title = {Hole spin precession and dephasing induced by nuclear hyperfine fields in ${\mathrm{CsPbBr}}_{3}$ and $\mathrm{CsPb}{(\mathrm{Cl},\mathrm{Br})}_{3}$ nanocrystals in a glass matrix},
  author = {Meliakov, Sergey R. and Belykh, Vasilii V. and Zhukov, Evgeny A. and Kolobkova, Elena V. and Kuznetsova, Maria S. and Bayer, Manfred and Yakovlev, Dmitri R.},
  journal = {Phys. Rev. B},
  volume = {110},
  issue = {23},

  year = {2024},

  doi = {10.1103/PhysRevB.110.235301},

}

@book{OpticalOrientation,
    author = {Meier, F and Zakharchenya, B. P} ,
    title = {Optical orientation},
    publisher = {Elsevier},
    year = {1984}
}

@book{SpinPhysicsinSC,
    author = {Dyakonov, M} ,
    title = {Spin Physics in Semiconductors},
    publisher = {Springer},
    year = {2017}
}

@article{NucSpinQD,
  title = {Nuclear spin physics in quantum dots: An optical investigation},
  author = {Urbaszek, Bernhard and Marie, Xavier and Amand, Thierry and Krebs, Olivier and Voisin, Paul and Maletinsky, Patrick and H\"ogele, Alexander and Imamoglu, Atac},
  journal = {Rev. Mod. Phys.},
  volume = {85},
  issue = {1},

  year = {2013},

  doi = {10.1103/RevModPhys.85.79},

}

@article{PRBMerkulov,
  title = {Electron spin relaxation by nuclei in semiconductor quantum dots},
  author = {Merkulov, I. A. and Efros, Al. L. and Rosen, M.},
  journal = {Phys. Rev. B},
  volume = {65},
  issue = {20},

  year = {2002},

  doi = {10.1103/PhysRevB.65.205309},

}

@article{HFcouplingconstants,
title = {Hyperfine coupling constants and atomic parameters for electron paramagnetic resonance data},
journal = {Atomic Data and Nuclear Data Tables},
volume = {33},
number = {2},

year = {1985},

doi = {https://doi.org/10.1016/0092-640X(85)90003-8},

author = {A.K. Koh and D.J. Miller},
}

@article{DFTNature,
  title = {Electronic structure of organometal halide perovskite CH3NH3BiI3 and optical absorption extending to infrared region},
  author = {Zhu, H. and Liu, JM},
  journal = {Scientific Reports},
  volume = {6},

  year = {2016},

  doi = {10.1038/srep37425},

}

@article{DFTexpLee,
doi = {10.1088/1361-6463/aa71e7},

year = {2017},

volume = {50},
number = {26},

author = {Lee, Min-I and Barragán, Ana and Nair, Maya N and Jacques, Vincent L R and Le Bolloc’h, David and Fertey, Pierre and Jemli, Khaoula and Lédée, Ferdinand and Trippé-Allard, Gaëlle and Deleporte, Emmanuelle and Taleb-Ibrahimi, Amina and Tejeda, Antonio},
title = {First determination of the valence band dispersion of CH3NH3PbI3 hybrid organic–inorganic perovskite},
journal = {Journal of Physics D: Applied Physics},

}

@article{DFTNatComm,
  title = {Absolute energy level positions in tin- and lead-based halide perovskites},
  author = {Tao, S. and Schmidt, I. and Brocks, G. et al},
  journal = {Nat Commun},
  volume = {10},

  year = {2019},

  doi = {10.1038/s41467-019-10468-7},

}

@article{OpticalOrientationTheory,
author = {Dyakonov, Michel and Perel, V.I.},
year = {1973},

title = {Optical orientation in a system of electrons and lattice nuclei in semiconductors. Theory},
volume = {65},
journal = {Journal of Experimental and Theoretical Physics}
}

@article{OpticalOrientationExperiment,
author = {Berkovits, Vladimir and Ekimov, Alexeï and Safarov, V},
year = {1973},

title = {Optical orientation in a system of electrons and lattice nuclei in semiconductors. Experiment},
volume = {65},
journal = {Journal of Experimental and Theoretical Physics}
}

@article{DPHyperfine,
author = {Dyakonov, Michel and Perel, V.},
year = {1973},

title = {Hyperfine Interaction in Optical Orientation of Electrons in Semiconductors},
volume = {36},
journal = {Journal of Experimental and Theoretical Physics - J EXP THEOR PHYS}
}

@article{DynamicKubo,
  title = {Zero-and low-field spin relaxation studied by positive muons},
  author = {Hayano, R. S. and Uemura, Y. J. and Imazato, J. and Nishida, N. and Yamazaki, T. and Kubo, R.},
  journal = {Phys. Rev. B},
  volume = {20},
  issue = {3},

  year = {1979},

  doi = {10.1103/PhysRevB.20.850},
}

@article{DTSCMDynamicKT,
doi = {10.1088/0031-8949/89/11/115201},

year = {2014},

volume = {89},
number = {11},

author = {Allodi, G and Renzi, R De},
title = {A numerical method to calculate the muon relaxation function in the presence of diffusion},
journal = {Physica Scripta},

}

@article{SCMKubo1954,
author = {Kubo, Ryogo},
title = {Note on the Stochastic Theory of Resonance Absorption},
journal = {Journal of the Physical Society of Japan},
volume = {9},
number = {6},
pages = {935-944},
year = {1954},
doi = {10.1143/JPSJ.9.935},
}

@article{SchultenWolynes,
    author = {Schulten, Klaus and Wolynes, Peter G.},
    title = {Semiclassical description of electron spin motion in radicals including the effect of electron hopping},
    journal = {The Journal of Chemical Physics},
    volume = {68},
    number = {7},

    year = {1978},

    issn = {0021-9606},
    doi = {10.1063/1.436135},

}

@book{Abragam,
      author        = "Abragam, Anatole",
      title         = "{The principles of nuclear magnetism}",
      publisher     = "Clarendon Press",

      year          = "1989",

}

@inproceedings{kubo1967magnetic,
  title={Magnetic resonance and relaxation},
  author={Kubo, Ryogo and Toyabe, T},
  booktitle={Proceedings of the XIVth colloque amp{\`e}re},
  volume={810},
  year={1967},
  organization={North-Holland}
}

@article{PRBKubo,
  title = {Zero-and low-field spin relaxation studied by positive muons},
  author = {Hayano, R. S. and Uemura, Y. J. and Imazato, J. and Nishida, N. and Yamazaki, T. and Kubo, R.},
  journal = {Phys. Rev. B},
  volume = {20},
  issue = {3},
  pages = {850--859},
  numpages = {0},
  year = {1979},
  month = {Aug},
  publisher = {American Physical Society},
  doi = {10.1103/PhysRevB.20.850},
  url = {https://link.aps.org/doi/10.1103/PhysRevB.20.850}
}

@article{TestelinHF,
  title = {Hole--spin dephasing time associated with hyperfine interaction in quantum dots},
  author = {Testelin, C. and Bernardot, F. and Eble, B. and Chamarro, M.},
  journal = {Phys. Rev. B},
  volume = {79},
  issue = {19},
 
  year = {2009},

  doi = {10.1103/PhysRevB.79.195440},

}

@article{SpinInertiaQD,
  title = {Theory of spin inertia in singly charged quantum dots},
  author = {Smirnov, D. S. and Zhukov, E. A. and Kirstein, E. and Yakovlev, D. R. and Reuter, D. and Wieck, A. D. and Bayer, M. and Greilich, A. and Glazov, M. M.},
  journal = {Phys. Rev. B},
  volume = {98},
  issue = {12},
  year = {2018},
  doi = {10.1103/PhysRevB.98.125306},

}

@article{SpinHoppingGlazov,
  title = {Spin noise of localized electrons: Interplay of hopping and hyperfine interaction},
  author = {Glazov, M. M.},
  journal = {Phys. Rev. B},
  volume = {91},


  year = {2015},


  doi = {10.1103/PhysRevB.91.195301},

}

@article{FAPbBrKirstein,
author = {Kirstein, Erik and Zhukov, Evgeny A. and Yakovlev, Dmitri R. and Kopteva, Nataliia E. and Yalcin, Ey{\"u}p and Akimov, Ilya A. and Hordiichuk, Oleh and Dirin, Dmitry N. and Kovalenko, Maksym V. and Bayer, Manfred},
title = {Coherent Carrier Spin Dynamics in FAPbBr3 Perovskite Crystals},
journal = {The Journal of Physical Chemistry Letters},
volume = {15},
number = {10},
pages = {2893-2903},
year = {2024},
doi = {10.1021/acs.jpclett.4c00098},
    note ={PMID: 38448798},

URL = { 
    
        https://doi.org/10.1021/acs.jpclett.4c00098
    
    

},
eprint = { 
    
        https://doi.org/10.1021/acs.jpclett.4c00098
    
    

}

}

@article{Polarrecovery,
  title = {Spin polarization recovery and Hanle effect for charge carriers interacting with nuclear spins in semiconductors},
  author = {Smirnov, D. S. and Zhukov, E. A. and Yakovlev, D. R. and Kirstein, E. and Bayer, M. and Greilich, A.},
  journal = {Phys. Rev. B},
  volume = {102},
  issue = {23},
  pages = {235413},
  numpages = {15},
  year = {2020},
  month = {Dec},
  publisher = {American Physical Society},
  doi = {10.1103/PhysRevB.102.235413},
  url = {https://link.aps.org/doi/10.1103/PhysRevB.102.235413}
}

@article{DNP2026BAYER,
  title = {Dynamic polarization of nuclear spins by optically oriented electrons and holes in lead halide perovskite semiconductors},
  author = {Kotur, Mladen and Bazhin, Pavel S. and Kavokin, Kirill V. and Kopteva, Nataliia E. and Yakovlev, Dmitri R. and Kudlacik, Dennis and Bayer, Manfred},
  journal = {Phys. Rev. B},
  volume = {113},
  issue = {8},
  pages = {085204},
  numpages = {22},
  year = {2026},
  month = {Feb},
  publisher = {American Physical Society},
  doi = {10.1103/w11v-2v4g},
  url = {https://link.aps.org/doi/10.1103/w11v-2v4g}
}

@Article{UnitCellVolumes,
author ="Baikie, Tom and Barrow, Nathan S. and Fang, Yanan and Keenan, Philip J. and Slater, Peter R. and Piltz, Ross O. and Gutmann, Matthias and Mhaisalkar, Subodh G. and White, Tim J.",
title  ="A combined single crystal neutron/X-ray diffraction and solid-state nuclear magnetic resonance study of the hybrid perovskites CH3NH3PbX3 (X = I{,} Br and Cl)",
journal  ="J. Mater. Chem. A",
year  ="2015",
volume  ="3",
issue  ="17",
pages  ="9298-9307",
publisher  ="The Royal Society of Chemistry",
doi  ="10.1039/C5TA01125F",
url  ="http://dx.doi.org/10.1039/C5TA01125F"
}

@article{UnitcellStoumpos,
author = {Stoumpos, Constantinos C. and Malliakas, Christos D. and Kanatzidis, Mercouri G.},
title = {Semiconducting Tin and Lead Iodide Perovskites with Organic Cations: Phase Transitions, High Mobilities, and Near-Infrared Photoluminescent Properties},
journal = {Inorganic Chemistry},
volume = {52},
number = {15},
pages = {9019-9038},
year = {2013},
doi = {10.1021/ic401215x}

}

@article{Gryncharova,

year = {1977},

volume = {11},

author = {Gryncharova, E.I and Perel, V.I},
title = {Relaxation of nuclear spin interacting with holes in semiconductors},
journal = {Sov. Phys. Semicond.},

}

@article{FAPILattice,
author = {Fabini, Douglas H. and Stoumpos, Constantinos C. and Laurita, Geneva and Kaltzoglou, Andreas and Kontos, Athanassios G. and Falaras, Polycarpos and Kanatzidis, Mercouri G. and Seshadri, Ram},
title = {Reentrant Structural and Optical Properties and Large Positive Thermal Expansion in Perovskite Formamidinium Lead Iodide},
journal = {Angewandte Chemie International Edition},
volume = {55},
number = {49},
pages = {15392-15396},
doi = {https://doi.org/10.1002/anie.201609538},
year = {2016}
}

@Article{MAPILattice,
author ="Lehmann, Frederike and Franz, Alexandra and Többens, Daniel M. and Levcenco, Sergej and Unold, Thomas and Taubert, Andreas and Schorr, Susan",
title  ="The phase diagram of a mixed halide (Br{,} I) hybrid perovskite obtained by synchrotron X-ray diffraction",
journal  ="RSC Adv.",
year  ="2019",
volume  ="9",
issue  ="20",
pages  ="11151-11159",
publisher  ="The Royal Society of Chemistry",
doi  ="10.1039/C8RA09398A",
}

@article{NanoLettersLAGUE,
author = {Lag{\"u}e, Guillaume and Bernardot, Fr{\'e}d{\'e}rick and Calvo, Mauricio and Testelin, Christophe and Miguez, Hernan and Chamarro, Maria},
title = {Hyperfine-Induced Asymmetric Overhauser Field in MAPbI3 Thin Films},
journal = {Nano Letters},
volume = {26},
number = {12},
pages = {4126-4133},
year = {2026},
doi = {10.1021/acs.nanolett.5c06420},

}

@article{Spintronics,
  title = {Spintronics: Fundamentals and applications},
  author = {\ifmmode \check{Z}\else \v{Z}\fi{}uti\ifmmode \acute{c}\else \'{c}\fi{}, Igor and Fabian, Jaroslav and Das Sarma, S.},
  journal = {Rev. Mod. Phys.},
  volume = {76},
  issue = {2},
  pages = {323--410},
  numpages = {0},
  year = {2004},
  month = {Apr},
  publisher = {American Physical Society},
  doi = {10.1103/RevModPhys.76.323},
  url = {https://link.aps.org/doi/10.1103/RevModPhys.76.323}
}

@article{ESpinDecoherenceQDs,
  title = {Electron Spin Decoherence in Quantum Dots due to Interaction with Nuclei},
  author = {Khaetskii, Alexander V. and Loss, Daniel and Glazman, Leonid},
  journal = {Phys. Rev. Lett.},
  volume = {88},
  issue = {18},
  pages = {186802},
  numpages = {4},
  year = {2002},
  month = {Apr},
  publisher = {American Physical Society},
  doi = {10.1103/PhysRevLett.88.186802},
  url = {https://link.aps.org/doi/10.1103/PhysRevLett.88.186802}
}

@article{SpinRelaxGaAs,
  title = {Low-temperature spin relaxation in n-type GaAs},
  author = {Dzhioev, R. I. and Kavokin, K. V. and Korenev, V. L. and Lazarev, M. V. and Meltser, B. Ya. and Stepanova, M. N. and Zakharchenya, B. P. and Gammon, D. and Katzer, D. S.},
  journal = {Phys. Rev. B},
  volume = {66},
  issue = {24},
  pages = {245204},
  numpages = {7},
  year = {2002},
  month = {Dec},
  publisher = {American Physical Society},
  doi = {10.1103/PhysRevB.66.245204},
  url = {https://link.aps.org/doi/10.1103/PhysRevB.66.245204}
}

@article{CdTeGuadalupe,
  title = {Spin relaxation time of donor-bound electrons in a CdTe quantum well},
  author = {Garcia-Arellano, G. and Bernardot, F. and Karczewski, G. and Testelin, C. and Chamarro, M.},
  journal = {Phys. Rev. B},
  volume = {99},
  issue = {23},
  pages = {235301},
  numpages = {10},
  year = {2019},
  month = {Jun},
  publisher = {American Physical Society},
  doi = {10.1103/PhysRevB.99.235301},
  url = {https://link.aps.org/doi/10.1103/PhysRevB.99.235301}
}

@article{InAsDesfonds,
    author = {Desfonds, P. and Eble, B. and Fras, F. and Testelin, C. and Bernardot, F. and Chamarro, M. and Urbaszek, B. and Amand, T. and Marie, X. and Gérard, J. M. and Thierry-Mieg, V. and Miard, A. and Lemaître, A.},
    title = {Electron and hole spin cooling efficiency in InAs quantum dots: The role of nuclear field},
    journal = {Applied Physics Letters},
    volume = {96},
    number = {17},
    pages = {172108},
    year = {2010},
    month = {04},
    issn = {0003-6951},
    doi = {10.1063/1.3394010},
    url = {https://doi.org/10.1063/1.3394010},
    eprint = {https://pubs.aip.org/aip/apl/article-pdf/doi/10.1063/1.3394010/14432131/172108_1_online.pdf},
}

@article{KudlacikFACsPbIBr,
author = {Kudlacik, Dennis and Kopteva, Nataliia E. and Kotur, Mladen and Yakovlev, Dmitri R. and Kavokin, Kirill V. and Harkort, Carolin and Karzel, Marek and Zhukov, Evgeny A. and Evers, Eiko and Belykh, Vasilii V. and Bayer, Manfred},
title = {Optical Spin Orientation of Localized Electrons and Holes Interacting with Nuclei in a FA0.9Cs0.1PbI2.8Br0.2 Perovskite Crystal},
journal = {ACS Photonics},
volume = {11},
number = {7},
pages = {2757-2769},
year = {2024},
doi = {10.1021/acsphotonics.4c00637},

URL = { 
    
        https://doi.org/10.1021/acsphotonics.4c00637
    
    

},
eprint = { 
    
        https://doi.org/10.1021/acsphotonics.4c00637
    
    

}

}

@article{PDOSCsPbBr,
doi = {10.1088/2053-1591/aca645},
url = {https://doi.org/10.1088/2053-1591/aca645},
year = {2022},
month = {dec},
publisher = {IOP Publishing},
volume = {9},
number = {12},
pages = {125501},
author = {Zaidi, S M Junaid and Khan, M Ijaz and Gillani, S S A and Sahar, M Sana Ullah and Ullah, Sana and Tanveer, Muhammad},
title = {A comprehensive DFT study to evaluate the modulation in the band gap, elastic, and optical behaviour of CsPbBr3 under the effect of stress},
journal = {Materials Research Express}
}

@Article{PDOSFAPI,
author ="Wang, Sanjun and Xiao, Wen-bo and Wang, Fei",
title  ="Structural{,} electronic{,} and optical properties of cubic formamidinium lead iodide perovskite: a first-principles investigation",
journal  ="RSC Adv.",
year  ="2020",
volume  ="10",
issue  ="54",
pages  ="32364-32369",
publisher  ="The Royal Society of Chemistry",
doi  ="10.1039/D0RA06028C"
}

@article{PDOSCsPbI,
author = {Gao, Li-Ke and Tang, Yan-Lin},
title = {Theoretical Study on the Carrier Mobility and Optical Properties of CsPbI3 by DFT},
journal = {ACS Omega},
volume = {6},
number = {17},
pages = {11545-11555},
year = {2021},
doi = {10.1021/acsomega.1c00734},
    note ={PMID: 34056310},

URL = { 
    
        https://doi.org/10.1021/acsomega.1c00734
    
    

},
eprint = { 
    
        https://doi.org/10.1021/acsomega.1c00734
    
    

}

}

@article{DFTMAPI,
author = {Feng, Jing and Xiao, Bing},
title = {Crystal Structures, Optical Properties, and Effective Mass Tensors of CH3NH3PbX3 (X = I and Br) Phases Predicted from HSE06},
journal = {The Journal of Physical Chemistry Letters},
volume = {5},
number = {7},
pages = {1278-1282},
year = {2014},
doi = {10.1021/jz500480m},
    note ={PMID: 26274484}

}

@article{Shumilin2015,
  title = {Kinetic equations for hopping transport and spin relaxation in a random magnetic field},
  author = {Shumilin, A. V. and Kabanov, V. V.},
  journal = {Phys. Rev. B},
  volume = {92},
  issue = {1},
  pages = {014206},
  numpages = {15},
  year = {2015},
  month = {Jul},
  publisher = {American Physical Society},
  doi = {10.1103/PhysRevB.92.014206},
  url = {https://link.aps.org/doi/10.1103/PhysRevB.92.014206}
}

@article{CLEMENTIROETTI,
title = {Roothaan-Hartree-Fock atomic wavefunctions: Basis functions and their coefficients for ground and certain excited states of neutral and ionized atoms, Z$\leq$54},
journal = {Atomic Data and Nuclear Data Tables},
volume = {14},
number = {3},
pages = {177-478},
year = {1974},
issn = {0092-640X},
doi = {https://doi.org/10.1016/S0092-640X(74)80016-1},
url = {https://www.sciencedirect.com/science/article/pii/S0092640X74800161},
author = {Enrico Clementi and Carla Roetti}
}

@article{HermanSkillman,
doi = {10.1149/1.2426131},
url = {https://doi.org/10.1149/1.2426131},
year = {1964},
month = {mar},
publisher = {The Electrochemical Society, Inc.},
volume = {111},
number = {3},
pages = {87C},
author = {Herman, Frank and Skillman, Sherwood and Arents, John},
title = {Atomic Structure Calculations},
journal = {Journal of The Electrochemical Society},

}

@article{MortonPreston,
title = {Atomic parameters for paramagnetic resonance data},
author = {J.R Morton and K.F Preston},
journal = {Journal of Magnetic Resonance (1969)},
volume = {30},

year = {1978},

doi = {https://doi.org/10.1016/0022-2364(78)90284-6},


}

@article{PRC2026Kotur,
  title = {Hyperfine interaction of electrons and holes with nuclei probed by optical orientation in methylammonium lead iodide perovskite crystals},
  author = {Kotur, Mladen and Kopteva, Nataliia E. and Yakovlev, Dmitri R. and Turedi, Bekir and Kovalenko, Maksym V. and Bayer, Manfred},
  journal = {Phys. Rev. B},
  volume = {113},
  issue = {24},
  pages = {245203},
  numpages = {9},
  year = {2026},
  month = {Jun},
  publisher = {American Physical Society},
  doi = {10.1103/3ptm-4tpd},
  url = {https://link.aps.org/doi/10.1103/3ptm-4tpd}
}

\end{document}